\documentclass[conference]{IEEEtran}
\IEEEoverridecommandlockouts
\usepackage[utf8]{inputenc} 
\usepackage[T1]{fontenc}    
\usepackage{hyperref}       
\usepackage{url}            
\usepackage{booktabs}       
\usepackage{amsfonts}       
\usepackage{nicefrac}       
\usepackage{microtype}      
\usepackage{xcolor}         

\usepackage{graphicx}
\usepackage{textcomp}
\usepackage{xcolor}
\usepackage{amssymb}

\usepackage{booktabs} 
\usepackage{float}
\usepackage{multirow}
\usepackage{indentfirst}
\usepackage{multirow}
\usepackage{amsmath}
\usepackage{graphicx}
\usepackage{subfigure}
\usepackage[]{graphicx,indentfirst}
\usepackage{arydshln}
\usepackage{textcomp}
\usepackage{multicol}
\usepackage{subfigure}
\usepackage{bm}
\usepackage{url}
\usepackage{amsfonts}
\usepackage{mathrsfs}

\usepackage{algorithmic}
\usepackage[ruled,linesnumbered]{algorithm2e}
\newcounter{chapter}

\newtheorem{definition}{Definition}[chapter]
\newtheorem{lemma}{Lemma}[chapter]
\newtheorem{proof}{Proof}[chapter]

\usepackage{ulem}
\usepackage{cite}
\usepackage{amsmath,amssymb,amsfonts}
\usepackage{algorithmic}
\usepackage{graphicx}
\usepackage{textcomp}
\usepackage{xcolor}
\usepackage[mathscr]{eucal}
\usepackage{arydshln}

\def\BibTeX{{\rm B\kern-.05em{\sc i\kern-.025em b}\kern-.08em
    T\kern-.1667em\lower.7ex\hbox{E}\kern-.125emX}}
\begin{document}

\title{Debiased Pairwise Learning from Positive-Unlabeled Implicit Feedback
}
\author{\IEEEauthorblockN{Bin Liu\IEEEauthorrefmark{2} , Qin Luo\IEEEauthorrefmark{2}  \thanks{\IEEEauthorrefmark{2}Bin Liu and Qin Luo contributed equally.}, Bang Wang\IEEEauthorrefmark{4} \thanks{ \IEEEauthorrefmark{4}Correspondence to: \url{wangbang@hust.edu.cn}.}}
\IEEEauthorblockA{\textit{School of Electronic Information and Communications}, \\
\textit{Huazhong University of Science and Technology (HUST)}, Wuhan, China,\\
Email: \{liubin0606, luo\_qin, wangbang\}@hust.edu.cn}
}
\maketitle

\begin{abstract}
Learning contrastive representations from pairwise comparisons has achieved remarkable success in various fields, such as natural language processing, computer vision, and information retrieval. Collaborative filtering algorithms based on pairwise learning also rooted in this paradigm. A significant concern is the absence of labels for negative instances in implicit feedback data, which often results in the random selected negative instances contains false negatives and inevitably, biased embeddings.  To address this issue, we introduce a novel correction method for sampling bias that yields a modified loss for pairwise learning called debiased pairwise loss (DPL). The key idea underlying DPL is to correct the biased probability estimates that result from false negatives, thereby correcting the gradients to approximate those of fully supervised data. The implementation of DPL only requires a small modification of the codes. Experimental studies on five public datasets validate the effectiveness of proposed learning method.
\end{abstract}


\section{Introduction}
Pairwise learning is a kind of learning paradigm that leverages pairwise comparisons to capture relative relationships between data pairs~\cite{gutmann:2012:JMLR,Ailon:2010:MachineLearning}. In contrast to pointwise learning, pairwise learning encourages an encoder to encode differential features between samples rather than pixel-level features of individual samples, usually leading to better generalization performance, particularly in scenarios where the absolute values of the samples are less meaningful~\cite{Wang:2020:ICML,McFadden:1974:FE,gutmann:2012:JMLR,Liu:2021:TKDE,Wang:2020:ICML,}. Pairwise learning has become a fundamental component of many modern machine learning algorithms and has facilitated significant advancements in various domains, including natural language processing, image and speech recognition, and recommendation systems~\cite{Oord:2018:arxiv,Wang:2020:ICML,He:2020:CVPR,Steffen:2009:UAI}. 

In the context of collaborative filtering, pairwise learning has been widely employed to predict rankings by contrasting positive and negative examples, with the most well-known approach being Bayesian Personalized Ranking (BPR)~\cite{Steffen:2009:UAI}, which has dominated the task of learning to rank  from  implicit feedback and achieved state-of-the-art performance. From a statistical perspective, BPR maximizes the posterior probability of observed ordered pairs between positive and negative samples. From a numerical computation perspective, the BPR loss function encourages the model to assign higher scores to positive examples than negative examples. In the embedding space, the BPR loss aims to pull the embedding of positive examples closer to the anchor embedding (i.e., user) while pushing negative examples apart from the anchor embedding. 

\begin{figure*}[h!]
	\centering
	\includegraphics[width=\textwidth]{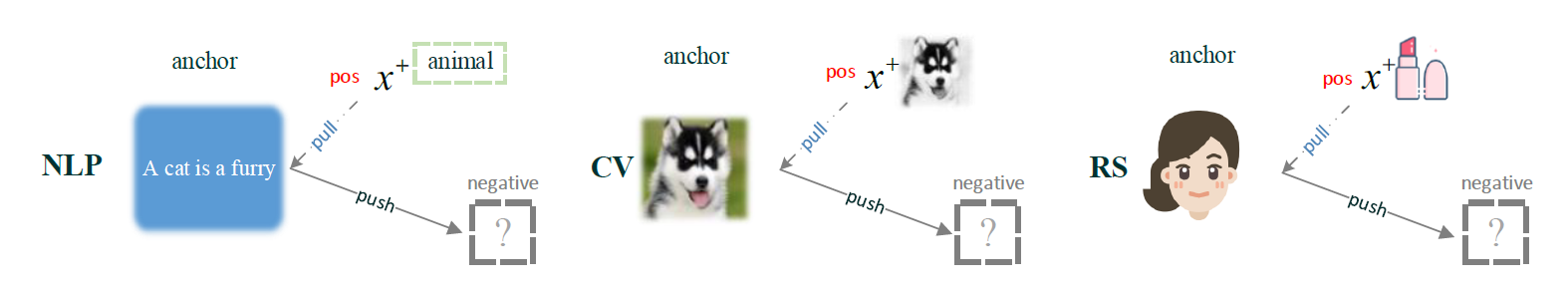}
	\caption{An illustrative example of the positive-unlabeled (PU) problem in machine learning. In natural language processing (NLP), positive samples can be obtained from the context, while negative examples are randomly sampled from vocabulary. In computer vision (CV), positive samples are obtained via data augmentation, while negative samples are sampled from unlabeled images. In recommendation systems (RS), positive samples are obtained from user interactions, while negative examples are randomly sampled from uninteracted items. Specifically, in recommendation systems, uninteracted items may be unseen item but be potentially liked in the future by the user, referred to as false negatives.}
	\label{Fig:falsen}
\end{figure*}

However, a prominent issue encountered in implicit collaborative filtering is obtaining negative feedback data can be challenging. This is mainly due to the fact that users usually only provide positive feedback by indicating their preferences or interests through interactions such as clicks, purchases, or ratings. As a result, the training set is often in the form of positive unlabeled (PU) data (Fig~\ref{Fig:falsen}), where only positive samples are available. PU data is commonly existed in various machine learning domains~\cite{Jessa:2020:ML,Du:2014:NIPS} such as unsupervised image classification~\cite{Chuang:2020:NIPS,Chen:2020:ICML,He:2020:CVPR,Chuang:2020:NIPS}, positive examples are obtained data augmentation. In practice, a common approach to handling positive-unlabeled implicit feedback is to treat un-interacted items as negative samples for model training~\cite{Steffen:2009:UAI,Xiangnan:2020:SIGIR,Wang:2019:SIGIR}. The resulting false negatives, i.e., items that user are preferred in future (positively labeled) but unseen during the training phase,  can significantly harm the performance of the model by introducing biases and incorrect representations of the users and items~\cite{Ding:2020:NIPS,Bin:2023:ICDE}.

To address the this issue, negative sampling has been widely investigated, and have shown promising results. Negative sampling can be classified into two types: the first kind is static negative sampling, which employs a fixed sampling distribution using some kind of side information that is independent of the model's training status, can result in easy samples, leading to suboptimal performance compared to dynamic negative sampling. Moreover, static negative sampling is severely limited by the availability of side information that serving as effective supervision signal. In contrast, dynamic negative sampling adjusts the sampling distribution using the model-dependent information such as predicted scores, aiming at sampling hard negative samples with high scores or top rankings to boost performance, which is prone to encounter false negative examples~\cite{Ding:2020:NIPS,Bin:2023:ICDE}. In addition, mini-batch training based on GPU batch computation require fixed positive and negative samples to be loaded into the dataloader before starting training, which may not be compatible with dynamic negative sampling. Dynamic negative sampling is typically implemented with additional computational and storage overhead, such as memorizing the predictive scores of previous training epochs or calculating the samples' predictive scores outside the mini-batch.

In this paper, we focus on the most general form of implicit feedback data, where there is no side information available for supervision. Specifically, we propose a correction for sampling bias from unlabeled data that yields a modified loss for pairwise learning called debiased pairwise loss (DPL). The key idea underlying DPL is to correct the biased probability estimates that result from false negatives, thereby correcting the gradients to approximate those of fully supervised data. The proposed objective is easy to implement and does not require additional side information for supervision or excessive storage and computational overhead.

\section{Preliminaries}
In this section, we introduce the notation, review Bayesian personalized ranking for implicit collaborative filtering.
\subsection{Notation}
Denote an user item pair $(u,i)$ as a sample $\mathbf x$, where $u\in \mathcal{U}, i\in \mathcal{I}$. Let $\mathcal{X}= \{\mathbf x|u\in \mathcal{U}, i\in \mathcal{I}\} $ be the sample space indicating all the user item pairs and $\mathcal{Y} =\{-1,+1\}$ be the class label indicating whether user prefer the item or not. A decision function $g:\mathcal{X} \rightarrow \mathbb{R}$ assigns value indicating the predicted preference level $g(\mathbf x) \in \mathbb{R}$. Denote the class conditional density of positives as $p^+(\mathbf x) = p(\mathbf x|+1)$, while class conditional density of positives as $p^-(\mathbf x) = p(\mathbf x|-1)$. So the marginal distribution $p(\mathbf x)=p^+(\mathbf x) \tau^+ +p^-(\mathbf x)\tau^- $, where $\tau^+ = 1-\tau^-$ is the prior probability $p(c(\mathbf{x}) = +1)$.
\subsection{Bayesian Personalized Ranking}
In implicit collaborative filtering, personalized ranking of a set of items are learned from pairwise comparisons of two randomly samples $(\mathbf{x}^+, \mathbf{x}^-)$. BPR\cite{Steffen:2009:UAI} adopts the well-known Bradley-Terry model to describe the likelihood of observing 
positive instance $\mathbf{x}^+$ been preferred over negative instances $\mathbf{x}^-$
\begin{eqnarray}
	\mathbb P(g(\mathbf{x}^+) > g(\mathbf{x}^-)|\Theta) = \sigma(g(\mathbf{x}^+) - g(\mathbf{x}^-))
\end{eqnarray}
where $\sigma(x) = \frac{1}{1+\exp(-x)}$ is the sigmoid function. BPR loss maximizes the probability of ordered pairs $g(\mathbf{x}^+) > g(\mathbf{x}^-)$ consisting of a positive instance and a negative instance:
\begin{eqnarray}\label{eq:bpr}
\mathcal{L}_{BPR} &=& - \mathbb{E}_{\substack{\mathbf x^+ \sim p^+(\mathbf x) \\ \mathbf x^- \sim p^-(\mathbf x^-)}} \log \sigma(g(\mathbf{x}^+) - g(\mathbf{x}^-)) \\
&=&  - \mathbb{E}_{\substack{\mathbf x^+ \sim p^+(\mathbf x) \\ \mathbf x^- \sim p^-(\mathbf x)}}\log \frac{1}{1+\exp(-g(\mathbf{x}^+) + g(\mathbf{x}^-))} \nonumber \\
&=&  - \mathbb{E}_{\substack{\mathbf x^+ \sim p^+(\mathbf x) \\ \mathbf x^- \sim p^-(\mathbf x^-)}}\log\frac{\exp(g(\mathbf{x}^+))}{\exp(g(\mathbf{x}^+))+\exp( g(\mathbf{x}^-))} \label{eq:infonce1}
\end{eqnarray}
It is worth noting that Eq. \eqref{eq:infonce1}, an equivalent form of the Bayesian Personalized Ranking (BPR) loss, is identical to the noise contrastive estimation (NCE) loss \cite{Gutmann:2010:ICAIS}, which is a special instance of the InfoNCE loss \cite{Oord:2018:arxiv} with a single negative sample (i.e., N=1). Moreover, in the collaborative filtering scenario, where users and items form a bipartite graph, typically only user embeddings are selected as anchor point.

In practice, positive instances are sampled from items that have been interacted with, denoted as $\mathbf{x}^+ \in \mathcal{D}^+$, while negative instances are sampled from items that have not been interacted with, denoted as $\mathbf{x}^- \in \mathcal{D}^-$.  Therefore, the empirical counterpart of Eq~\eqref{eq:bpr} is given by:
\begin{eqnarray}\label{eq:bpr_emp}
\mathcal{L}_\text{BPR} =- \frac{1}{|\mathcal{D}^+|\times |\mathcal{D}^-|} \sum_{\mathbf{x}^+ \in \mathcal{D}^+}\sum_{\mathbf{x}^- \in \mathcal{D}^-} && \ln \sigma(g(\mathbf{x}^+) - g(\mathbf{x}^-)) \nonumber \\&& - \lambda ||\Theta||^2,
\end{eqnarray}
where $\lambda ||\Theta||^2$ is the regularization term to balance the variance and bias and avoid over-fitting. Notably, the regularization term $\lambda ||\Theta||^2$ is equivalent to the logarithm of the prior density of a Gaussian distribution, thereby offering a posterior probability-based interpretation to Eq~\eqref{eq:bpr_emp}. In the original Bayesian Personalized Ranking (BPR) paper~\cite{Steffen:2009:UAI}, Eq~\eqref{eq:bpr_emp} is interpreted as the maximum posterior estimator of observed ordered pairs.
\section{Proposed Method}
\subsection{Sampling bias}
Due to the absence of labeled negative samples, one can only sample negative examples from unlabeled data during training for optimizing Eq~\eqref{eq:bpr_emp}, resulting in the following biased optimization objective:
\begin{eqnarray}\label{eq:bpr_emp_biased}
	\mathcal{L} =- \frac{1}{|\mathcal{D}^+|\times |\mathcal{D}^u|} \sum_{\mathbf{x}^+ \in \mathcal{D}^+}\sum_{\mathbf{x} \in \mathcal{D}^u} && \ln \sigma(g(\mathbf{x}^+) - g(\mathbf{x})) \nonumber \\&& - \lambda ||\Theta||^2,
\end{eqnarray}
Next, we investigate the impact of biased optimization objectives on the learned embeddings of user-item pairs $\Theta$:
\begin{eqnarray}
\frac{\partial \mathcal{L}}{\partial \Theta} &=& \frac{\partial \mathcal{L}_\text{BPR}}{g(\mathbf{x})}\cdot\frac{g(\mathbf{x})}{\Theta}  \\
&=& [1-\sigma(g(\mathbf{x}^+) - g(\mathbf{x})) ]\cdot\frac{g(\mathbf{x})}{\Theta}  \label{eq:graident}
\end{eqnarray}
Eq~\eqref{eq:graident} is the result of the differential chain rule, where the first term $[1-\sigma(g(\mathbf{x}^+) - g(\mathbf{x})) ]$ is determined by the form of the loss function and the second term $\frac{g(\mathbf{x})}{\Theta}$ is determined by the decision function. For a fixed model, the second term remains the same. 

\begin{figure}[h!]
	\centering
	\includegraphics[width=0.5\textwidth]{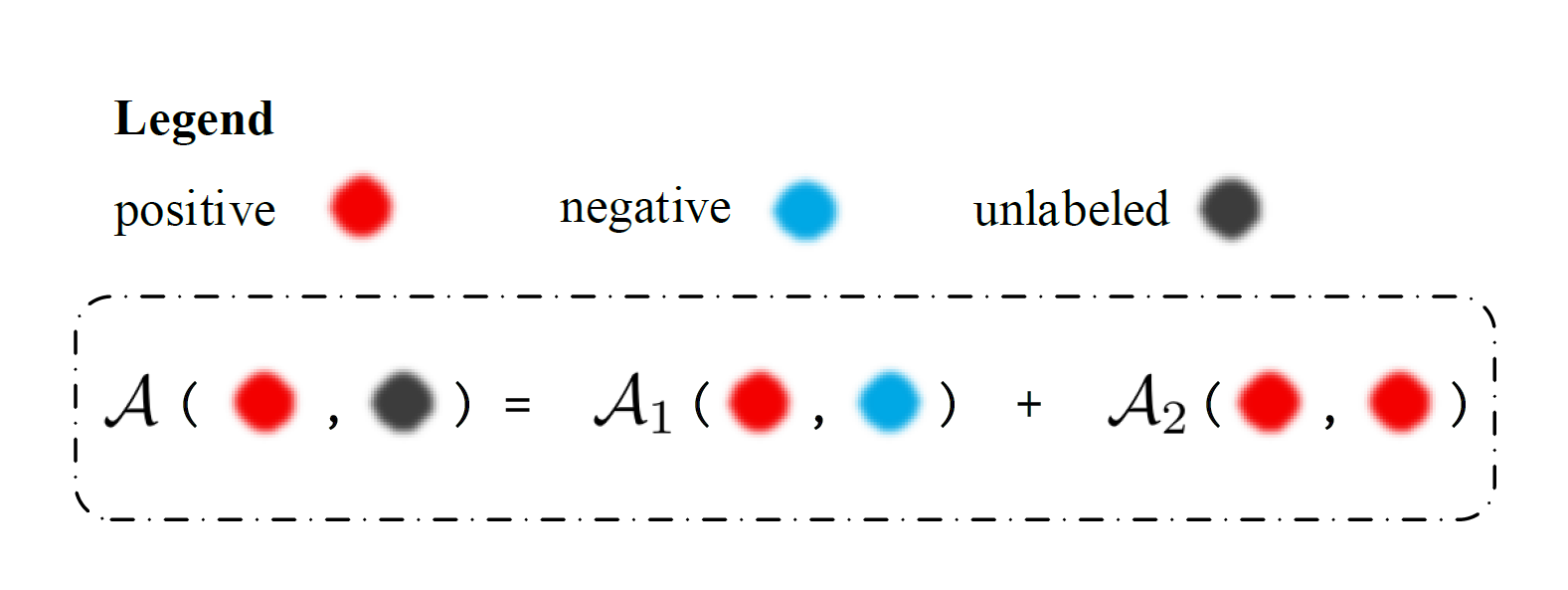}
	\caption{An illustrative example that explains the value $\sigma(g(\mathbf{x}^+) - g(\mathbf{x}))$ calculate using positive-unlabeled data pairs is a biased probability estimate. Based on the principle of inclusion-exclusion, the event $\mathcal{A}(+,u) = \mathcal{A}_1(+,-) + \mathcal{A}_2(+,+)$. Note the objective is to maximize the likelihood of positive item being preferred over negative item, so the probability, which models the event $\mathcal{A}_1(+,-)$ consisting of positive-negative pairs, is the optimization objective. However, in practice we optimize the biased probability that models the event $\mathcal{A}(+,u)$.} 
	\label{fig:event}
\end{figure}

In the first term of Eq~\eqref{eq:graident}, the real-valued sigmoid function $\sigma(g(\mathbf{x}^+) - g(\mathbf{x})) \in [0,1]$ is been interpreted as the likelihood of positive item $\mathbf{x}^+$ being preferred over negative item~\cite{Steffen:2009:UAI}. However, since $\mathbf{x}$ is an unlabeled sample with a positive class prior $\tau^+$, this leads to a biased estimate of the probability (see Fig~\ref{fig:event}). Intuitively, biased $\sigma(g(\mathbf{x}^+) - g(\mathbf{x}))$-values will lead to incorrect gradient magnitude ${\partial \mathcal{L}_\text{BPR}}/{\partial \Theta}$, resulting in inaccurate user-item representations when performing stochastic gradient descent learning algorithm.

To maintain notational brevity in mathematical expressions, we define a mapping $h: \mathcal{X}\times\mathcal{X} \rightarrow \sigma(g(\mathbf{x}^+) - g(\mathbf{x}))$ that maps two samples into a sigmoid function. The problem then becomes how to approximate the value of $h(\mathbf{x}^+,\mathbf{x}^-)$ using samples from the positive and unlabeled populations, respectively. Specifically, given a set of positive samples $\{\mathbf{x}^+_i\}_{i=1}^M$ and a set of unlabeled samples $\{ \mathbf{x}_j\}_{j=1}^N$, we aim to estimate the value of $h(\mathbf{x}^+,\mathbf{x}^-)$. By doing so, we can correct the gradients to approximate those of fully supervised data, thereby leading to better generalization performance of learned user-item representations.

\subsection{Bias Correction}
In order to approximate the value of $h(\mathbf{x}^+,\mathbf{x}^-)$ using a set of positive samples and a set of unlabeled samples, we start by establishing a relationship between the joint distribution of positive-unlabeled sample pairs, denoted by $p_\textsc{pu}$, and the joint distribution of positive-negative sample pairs, denoted by $p_\textsc{pn}$.

The first sample of a sample pair, denoted by $\mathbf x_1$, is deterministically drawn from the positive class conditional probability $p^+(\mathbf x)$, while the second sample, denoted by $\mathbf{x}_2$, is drawn from the marginal distribution $p(\mathbf{x})$. Consequently, the joint distribution of positive-unlabeled sample pairs $p_\textsc{pu}$ can be expressed as follows:
\begin{eqnarray}
	p_\textsc{pu}(\mathbf{x}_1, \mathbf{x}_2) &=& p^+( \mathbf{x}_1) p( \mathbf{x}_2)  \label{eq:independent} \\
	&=& p^+( \mathbf{x}_1) [p^+(\mathbf x_2) \tau^+ +p^-(\mathbf x_2)\tau^- ] \label{eq:full}\\
	&=& \tau^+p^+( \mathbf{x}_1) p^+(\mathbf x_2)  \nonumber \\
	&&+ \tau^-p^+( \mathbf{x}_1)p^-(\mathbf x_2) \label{eq:pnpp}
\end{eqnarray}
Eq.~\eqref{eq:independent} is obtained since $(\mathbf{x}_1, \mathbf{x}_2)$ are independently drawn.Meanwhile, Eq.~\eqref{eq:full} is the full probability decomposition of the probability of the marginal distribution $p(\mathbf{x})$. By rearranging Eq.~\eqref{eq:pnpp}, we can establish a relationship between the desired joint distribution $p_\textsc{pn}(\mathbf{x}_1, \mathbf{x}2)$ of positive-negative sample pairs and the joint distribution $p_{\textsc{pu}}(\mathbf{x}_1, \mathbf{x}_2)$ of positive-unlabeled sample pairs, expressed as:
\begin{eqnarray}\label{eq:jointpn}
	p_\textsc{pn}(\mathbf{x}_1, \mathbf{x}_2)  &=& p^+( \mathbf{x}_1)p^-(\mathbf x_2) \nonumber \\
	&=& \frac{1}{\tau^-}[p_{\textsc{pu}}(\mathbf{x}_1, \mathbf{x}_2)- \tau^+p^+( \mathbf{x}_1) p^+(\mathbf x_2)] \label{eq:pndist}
\end{eqnarray}
Therefore, the expected value of $h(\mathbf x_1,\mathbf x_2)$ over the desired joint distribution $p_\textsc{pn}(\mathbf{x}_1, \mathbf{x}_2)$ of positive-negative sample pairs, which represents the expected likelihood of a positive sample $\mathbf{x}_1$ being preferred over $\mathbf{x}_2$ under fully labeled data, can be computed as follows:
\begin{eqnarray}
	&&\mathbb{E}_{ p_\textsc{pn}(\mathbf{x}_1, \mathbf{x}_2)} h(\mathbf{x}_1,\mathbf{x}_2)\\
	&=& \int_{\mathbf{x}_1}\int_{\mathbf{x}_2}   h(\mathbf{x}_1,\mathbf{x}_2) p_\textsc{pn}(\mathbf{x}_1, \mathbf{x}_2) d{\mathbf{x}_1}d{\mathbf{x}_2} \\
&=&\int_{\mathbf{x}_1}\int_{\mathbf{x}_2}   h(\mathbf{x}_1,\mathbf{x}_2) [\frac{1}{\tau^-}p_{\textsc{pu}}(\mathbf{x}_1, \mathbf{x}_2)\nonumber \\
&&- \frac{\tau^+}{\tau^-}p^+( \mathbf{x}_1) p^+(\mathbf x_2)]d{\mathbf{x}_1}d{\mathbf{x}_2} \\
&=&\int_{\mathbf{x}_1}\int_{\mathbf{x}_2}   h(\mathbf{x}_1,\mathbf{x}_2) [\frac{1}{\tau^-}p_{\textsc{pu}}(\mathbf{x}_1, \mathbf{x}_2) d{\mathbf{x}_1}d{\mathbf{x}_2}\nonumber \\
&&- \int_{\mathbf{x}_1}\int_{\mathbf{x}_2} \frac{\tau^+}{\tau^-}p^+( \mathbf{x}_1) p^+(\mathbf x_2)]d{\mathbf{x}_1}d{\mathbf{x}_2} \\
&=& \frac{1}{\tau^-}\mathbb{E}_{ p_\textsc{pu}(\mathbf{x}_1, \mathbf{x}_2)} h(\mathbf{x}_1,\mathbf{x}_2) -\frac{\tau^+}{\tau^-} \mathbb{E}_{ p_\textsc{pp}(\mathbf{x}_1, \mathbf{x}_2)} h(\mathbf{x}_1,\mathbf{x}_2) \label{eq:expec}
\end{eqnarray}
\begin{algorithm*}[!h]\label{Alg:1}
	\normalem
	\small
	\caption{Pseudocode of proposed Debiased Pairwise Learning (DPL) method in a PyTorch-like style.}
	\label{Alg}
	\KwIn{Mini-batch data $\mathcal{R}$, number of extra positive items $M$, number of negative items $N$, positive class prior \textit{tau}, batch size \textit{bs}, score function $g(\cdot)$ ($M,N \geq 1$).}
	\KwOut{Embeddings $\Theta \in \mathbb{R}^d$}
	\verb|scores     = | $g(\mathcal{R})$\verb|  #[bs*(1+M+N)] the scores of each items in the mini-batch data |\\
	\verb|pos_scores = scores[:,:M+2] |~~~~~~~~~~~~~~~~~~~~~~~~~~~~~~~~~~~~~~~~~~~~~~~~~~~~~~~~~~~~~~~~~~~~~~~~~~~~~~~~~~~\verb|#[bs*(1+M)]| \\
	\verb|neg_scores = scores[:,M+2:]                       |~~~~~~~~~~~~~~~~~~~~~~~~~~~~~~~~~~~~~~~~~~~~~~\verb|#[bs*N]| \\
	\verb|pu_prob    = sigmoid(pos_scores[:, 0:1] - neg_scores)                      |    ~~~~\verb|#[bs*N]| \\
	\verb|pp_prob    = sigmoid(pos_scores[:, 0:1] - pos_scores[:,1:])                |    ~~~~\verb|#[bs*M]| \\
	\verb|pn_prob    = pu_prob.mean(dim=-1)/(1-tau) - tau*pp_prob.mean(dim=-1)/(1-tau)|~~~\verb|#[bs, ]| \\
	\verb|dpl_loss   = - log (pn_prob).mean()   |~~~~~~~~~~~~~~~~~~~~~~~~~~~~~~~~~~~~~~~~~~~~~~~~~~~~~~~~~~~~~~~~~~~~~~~\\
	\verb|Update embeddings based on gradient w.r.t. dpl_loss.|\\
	\KwResult{Final embeddings.}
\end{algorithm*}
The first term of Eq.\eqref{eq:expec} represents the expectation over positive-unlabeled sample pairs, and this term can be estimated empirically using $N$ unlabeled samples $\{\mathbf{x}_n\}_{n=1}^N$ given the first positive sample $\mathbf{x}^+$:
\begin{eqnarray}
	\mathbb{E}_{ p_\textsc{pu}(\mathbf{x}_1, \mathbf{x}_2)} h(\mathbf{x}_1,\mathbf{x}_2) &=& \frac{1}{N} \sum_{n=1}^{N} h(\mathbf{x}^+,\mathbf{x}_n)\label{eq:epu},
\end{eqnarray}
and the second term of Eq.\eqref{eq:expec} represents the expectation over positive-positive sample pairs. This term can be estimated empirically using $M$ additional positive samples $\{\mathbf{x}^\prime_n\}_{m=1}^M$ given the first positive sample $\mathbf{x}^+$:
\begin{eqnarray}
\mathbb{E}_{ p_\textsc{pp}(\mathbf{x}_1, \mathbf{x}_2)} h(\mathbf{x}_1,\mathbf{x}_2)  &=& \frac{1}{M} \sum_{m=1}^{M} h(\mathbf{x}^+,\mathbf{x}_m^{\prime})\label{eq:epp}.
\end{eqnarray}
Inserting Eq.~\eqref{eq:epu} and Eq.~\eqref{eq:epp} back to Eq.~\eqref{eq:expec}, we obtain the final empirical estimate:
\begin{eqnarray}\label{eq:est}
	\hat{P}_\textsc{pn} = \frac{1}{N\tau^-}
	\sum_{\mathbf{x} \in \mathcal{D}^u}h(\mathbf{x}^+,\mathbf{x})  - \frac{\tau^+}{M\tau^-}\sum_{\mathbf{x}^\prime\in \mathcal{D}^+}h(\mathbf{x}^+, \mathbf{x}^\prime)
\end{eqnarray}
As shown, the first term in the equation is computed using positive-unlabeled sample pairs, but with an additional correction term included to offset the biased likelihood estimation resulting from false negative samples being included in the unlabeled set. The correction term computed from positive-positive pairs in Eq. \eqref{eq:est} may appear peculiar, and we refer the reader to Fig. \ref{fig:auc} for an intuitive explanation of this correction term. The final empirical form of the debiased pairwise loss (DPL) is presented below:
\begin{eqnarray}\label{eq:dpl}
\mathcal{L}_\textsc{dpl}=- \frac{1}{|\mathcal{D}^+|\times |\mathcal{D}^u|} \sum_{\mathbf{x}^+ \in \mathcal{D}^+}\sum_{\mathbf{x} \in \mathcal{D}^u} && \ln \hat{P}_\textsc{pn} \nonumber.
\end{eqnarray}

\subsection{Implementation}
When training with the BPR loss, a negative sample $j$ is required for each $(u,i)$ pair, resulting in a data entry represented as a $(u,i,j)$ triplet. To address sampling bias, the DPL loss requires $M\geq 1$ additional positive examples and $N\geq1$ negative examples for each $(u,i)$ pair. Following the same data entry format as BPR, each DPL data entry is organized as $(u,i,i_1,i_2,\cdots,i_M,j_1,j_2,\cdots,j_N)$. This can be easily implemented by rewriting the \verb|collate_fn| function of the \verb|Dataloader|. As a result, each mini-batch data is structured as follows:
\begin{eqnarray}\nonumber
\text{batch size}\left\{ \left[\begin{array}{cc:cccc:cccc}
u^{1} & {i}   &i_{1} & i_{2} & \ldots &i_{M} &   j_{1}& j_{2} & \ldots &j_{N} \\
u^{2} & i & i_{1}& i_{2} & \ldots &i_{M} &  j_{1}& j_{2} & \ldots &j_{N} \\
\vdots & \vdots & \vdots & \vdots &\ddots & \vdots & \vdots & \vdots & \ddots & \vdots \\
u^{bs} & i & i_{1}& i_{2} & \ldots &i_{M} &  j_{1}& j_{2} & \ldots &j_{N} 
\end{array}\right]\right.
\end{eqnarray}
Each data entry includes N unlabeled items $j_1,j_2,\cdots,j_N$, and N predicted scores $\hat{x}_{j}^1,\hat{x}_{j}^2,\cdots,\hat{x}_{j}^N$ for these items can be calculated. For the given positive sample $(u,i)$ with its score $\hat{x}_i$, N pu probability values can be obtained based on the predicted scores of the N negative samples. Therefore, the estimated PU probability value based on the mini-batch data is given by:
\[\hat{P}_\textsc{pu} = \frac{1}{N} \sum_{n=1}^{N}\sigma (\hat{x}_i - \hat{x}_{j}^n)\]
Similarly, M predicted scores $\hat{x}_{i}^1,\hat{x}_{i}^2,\cdots,\hat{x}_{i}^M$ for M positive items $i_1,i_2,\cdots,i_M$ can be computed, and the estimated PP probability value based on the mini-batch data is given as follows:
\[\hat{P}_\textsc{pp} = \frac{1}{M} \sum_{m=1}^{M}\sigma (\hat{x}_i - \hat{x}_{i}^m)\]
Therefore, the corrected probability value that approximates the probability of user preferred positive item over negative item is given by:
\[\hat{P}_\textsc{pn} = \frac{1}{\tau^-}
\hat{P}_\textsc{pu} - \frac{\tau^+}{\tau^-}\hat{P}_\textsc{pp} \]
Algorithm~\ref{Alg} presents the pseudocode for the DPL algorithm in a PyTorch-like style.

\textbf{Complexity}: We first analyze the complexity of the baseline method BPR, which is related to the scoring function $g$. Here, we take matrix factorization with latent dimension $d$ as an example, and this analysis can be easily extended to other models. Given a mini-batch data with a batch size of $bs$ consisting of $(u,i,j)$ training triples, the forward scoring calculation involves a total of $2\times bs$ item score predictions, resulting in a time complexity of $\mathcal O(2 bs\times d)$. In backward propagation, at most $3 \times bs$ embeddings are updated, and a total of $5bs\times d$ operations are involved, resulting in a time complexity of $\mathcal O(bs\times d)$. Similarly, for DPL, a mini-batch data involves a total of $(M+N+1)\times bs$ scoring calculations and $(M+N+2)\times bs$ embedding updates, involving a total of $(2M+2N+3)\times bs\times d$ operations and a time complexity of $\mathcal O(bs\times d)$, since $M$ and $N$ are usually set to small constants such as $M=3$ and $N=3$ in practice. Specifically, when $M=0$ and $N=1$, the number of operations involved in DPL is the same as that in BPR. Therefore, DPL has strictly linear complexity relative to BPR, without any calculation or storage overhead outside of mini-batch data.

\section{Theoretical Analysis}
\begin{figure*}[h!]
	\centering
	\includegraphics[width=\textwidth]{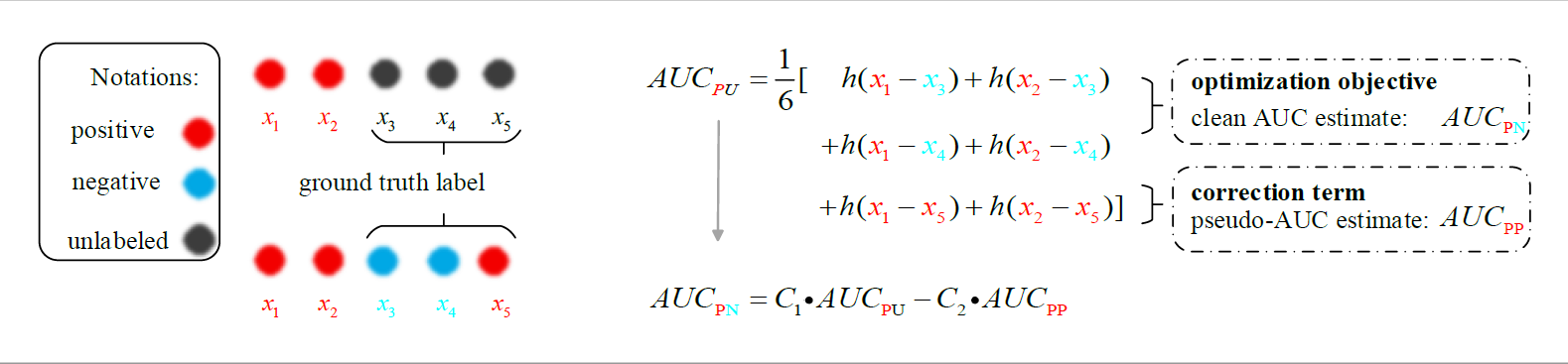}
	\caption{An illustrative example that explains the origin of the correction term involving two positive samples. Let us consider a positive-unlabeled dataset, where ${x_1,x_2}$ are positive samples and ${x_3,x_4,x_5}$ are unlabeled samples. The ground truth labels of the unlabeled samples are not accessible during training. The area under the curve (AUC) that is computed using positive examples and unlabeled data pairs is denoted as $AUC_\textsc{pu}$. It can be seen that $AUC_\textsc{pu}$ is the sum of two terms. The first term is the clean AUC estimate computed using positive-negative (PN) data pairs and is denoted as $AUC_\textsc{pn}$. The second term is the pseudo AUC estimate computed using positive-positive (PP) data pairs and is denoted as $AUC_\textsc{pp}$. To optimize the clean AUC, we should subtract the $AUC_\textsc{pp}$ from the $AUC_\textsc{pu}$.} 
	\label{fig:auc}
\end{figure*}

The main idea behind DPL is to improve the biased probability that calculated using the positive-unlabeled data pairs. This is achieved by sampling additional positive and negative examples to estimate the expected probability value of users liking positive items more than negative items, which is given by Eq. \eqref{eq:est}, and used to replace the original biased probability estimate. To demonstrate Eq. \eqref{eq:est} is a good estimator, we first prove that it is an unbiased estimator of the AUC risk.
\begin{lemma}\label{lemma:auc} Let positive data $\mathbf{x}^+$ i.i.d. drawn from positive class conditional density $p^+(\mathbf{x})$, and unlabeled data  $\mathbf{x}$ i.i.d. drawn from marginal density $p(\mathbf{x}) = \tau^+ p^+(\mathbf{x})+ \tau^- p^-(\mathbf{x})$. Then Eq.~\eqref{eq:est} is the unbiased estimate of AUC risk:
	\[\mathbb{E}\hat{Pr} = R_{AUC}\]
	\begin{proof}
\begin{eqnarray}
\mathbb{E}\hat{Pr} &=& \int_{\mathbf{x}^+} [\frac{1}{N\tau^-}
\sum_{\mathbf{x} \in \mathcal{D}^u} \mathbb{E}_{\mathbf{x}\sim p}h(\mathbf{x}^+,\mathbf{x}) \nonumber \\ &&- \frac{\tau^+}{M\tau^-}\sum_{\mathbf{x}^\prime\in \mathcal{D}^+}\mathbb{E}_{\mathbf{x}^\prime \sim p^+}h(\mathbf{x}^+, \mathbf{x}^\prime)]p^+(\mathbf{x}^+)d\mathbf{x}^+ \\
&=&\int_{\mathbf{x}^+} [\frac{1}{\tau^-}
\mathbb{E}_{\mathbf{x}\sim p}h(\mathbf{x}^+,\mathbf{x}) \nonumber \\ &&- \frac{\tau^+}{\tau^-}\mathbb{E}_{\mathbf{x}^\prime \sim p^+}h(\mathbf{x}^+, \mathbf{x}^\prime)]p^+(\mathbf{x}^+)d\mathbf{x}^+ \label{eq:unbiasauc}.
\end{eqnarray}
Since
\begin{eqnarray}
	&&\frac{1}{\tau^-}\mathbb{E}_{\mathbf{x}\sim p}h(\mathbf{x}^+,\mathbf{x}) - \frac{\tau^+}{\tau^-}\mathbb{E}_{\mathbf{x}^\prime\sim p^+}h(\mathbf{x}^+, \mathbf{x}^\prime)\nonumber \\
	&=&\frac{1}{\tau^-} \int_\mathbf{x}h(\mathbf{x}^+,\mathbf{x})p(\mathbf{x})d\mathbf{x} - \frac{\tau^+}{\tau^-} \int_\mathbf{x^\prime}h(\mathbf{x}^+, \mathbf{x}^\prime)p^+(\mathbf{x}^\prime)d\mathbf{x^\prime} \nonumber \\
	&=&\frac{1}{\tau^-} \int_\mathbf{x}h(\mathbf{x}^+,\mathbf{x})[\tau^+p^+(\mathbf{x}) + \tau^-p^-(\mathbf{x}) ]d\mathbf{x} \label{eq:marginal}
	\\&&- \frac{\tau^+}{\tau^-} \int_\mathbf{x^\prime}h(\mathbf{x}^+, \mathbf{x}^\prime)p^+(\mathbf{x}^\prime)d\mathbf{x^\prime}  \nonumber \\
	&=& \int_\mathbf{x}h(\mathbf{x}^+,\mathbf{x})p^-(\mathbf{x})d\mathbf{x} \label{eq:variable} \\
	&=& \int_\mathbf{x^-}h(\mathbf{x}^+,\mathbf{x}^-)p^-(\mathbf{x}^-)d\mathbf{x}^- \label{eq:pmin} 
\end{eqnarray}
where Eq~\eqref{eq:marginal} is obtained by decomposing the marginal distribution $p(\mathbf{x})= \tau^+p^+(\mathbf{x}) + \tau^-p^-(\mathbf{x})$, and Eq~\eqref{eq:pmin} replaces the integration variable $\mathbf{x}$ in Eq~\eqref{eq:variable} with $\mathbf{x}^-$ to enhance readability. Inserting Eq~\eqref{eq:pmin} back to Eq~\eqref{eq:unbiasauc} we obtain
\begin{eqnarray}
	\mathbb{E}\hat{P}_r &=& \int_{\mathbf{x}^+} \int_\mathbf{x}h(\mathbf{x}^+,\mathbf{x}^-)p^-(\mathbf{x}^-) p^+(\mathbf{x}^+)d\mathbf{x}^+d\mathbf{x}^- \label{eq:aucrisk}\\
	&=& R_{AUC}.
\end{eqnarray}

which completes the proof. If we substitute the function $h$ in Eq~\eqref{eq:aucrisk} with the 0-1 loss $\mathbb{I}(\mathbf{x}^+,\mathbf{x}^-)$, then Eq~\eqref{eq:aucrisk} exactly defines the AUC metric. However, due to the discrete nature of the 0-1 loss function, a differentiable surrogate loss is often used in practice when optimizing for AUC metric. We refer to Fig.~\ref{fig:auc} for an intuitive explanation of Lemma~\ref{lemma:auc}.
\end{proof}	
\end{lemma}

Next, we seek a  an idealized objective for DPL to approximate.
\begin{definition}
For fixed positive sample $\mathbf{x}^+$, let $\mathbb{P}_\textsc{pn} = \mathbb{E}_{\mathbf{x}^-\sim p^-}h(\mathbf{x}^+,\mathbf{x}^-)$ being the expected probability value of positive sample $\mathbf{x}^+$ being preferred over true negative item $\mathbf{x}^-$. Then, taking the log-likelihood over all positive samples, we can define the supervised loss as:
\begin{eqnarray}
\mathcal{L}_\textsc{sup} =  -\mathbb{E}_{\mathbf{x}^+\sim p^+} \log \mathbb{P}_\textsc{pn}
\end{eqnarray}
\end{definition}
Minimizing the expected log-likelihood $-\log \mathbb{P}_\text{PN}$ over all positive samples results in the maximization of the likelihood that any positive sample is preferred over any true negative sample, which is exactly our optimization objective under fully supervised data. Therefore, $\mathcal{L}_\textsc{sup}$ is an idealized objective for DPL to approximate. Lemma~\ref{lemma:asy} proves that the DPL estimate is asymptotically consistent with this idealized loss as M and N approach infinity.
\begin{lemma}\label{lemma:asy}
For $M,N \rightarrow +\infty$, we have
\begin{eqnarray}
\mathcal{L}_\textsc{dpl}  \rightarrow \mathcal{L}_\textsc{sup} 
\end{eqnarray}
\begin{proof}
Lebesgue's dominated converge theorem states that for a sequence of measurable functions $f_n$ that are bounded, then
\begin{eqnarray}
\lim\limits_{n\rightarrow \infty} \int_{\Omega} f_n =\int_{\Omega} \lim\limits_{n\rightarrow\infty}f_n \nonumber
\end{eqnarray}
so
\begin{eqnarray}
&&\lim\limits_{\substack{M,N \rightarrow +\infty}} \mathcal{L}_\textsc{dpl} \nonumber\\
&=&-\lim\limits_{\substack{M,N \rightarrow +\infty}} \mathbb{E}_{\mathbf{x}^+\sim p^+}\log [\frac{1}{N\tau^-}
\sum_{\mathbf{x} \in \mathcal{D}^u}h(\mathbf{x}^+,\mathbf{x})  \\&&~~~~~~~~~~- \frac{\tau^+}{M\tau^-}\sum_{\mathbf{x}^\prime\in \mathcal{D}^+}h(\mathbf{x}^+, \mathbf{x}^\prime)]\nonumber \\
&=&-\mathbb{E}_{\mathbf{x}^+\sim p^+}\lim\limits_{\substack{M,N \rightarrow +\infty}} \log [\frac{1}{N\tau^-}
\sum_{\mathbf{x} \in \mathcal{D}^u}h(\mathbf{x}^+,\mathbf{x}) \\&&~~~~~~~~~~ - \frac{\tau^+}{M\tau^-}\sum_{\mathbf{x}^\prime\in \mathcal{D}^+}h(\mathbf{x}^+, \mathbf{x}^\prime)]\nonumber \\
&=&-\mathbb{E}_{\mathbf{x}^+\sim p^+}\lim\limits_{\substack{M,N \rightarrow +\infty}} \log [\frac{1}{N\tau^-}
\sum_{\mathbf{x} \in \mathcal{D}^u}h(\mathbf{x}^+,\mathbf{x}) \\&&~~~~~~~~~~ - \frac{\tau^+}{M\tau^-}\sum_{\mathbf{x}^\prime\in \mathcal{D}^+}h(\mathbf{x}^+, \mathbf{x}^\prime)]\nonumber \\
&=&-\mathbb{E}_{\mathbf{x}^+\sim p^+}\log [\frac{1}{\tau^-}\mathbb{E}_{\mathbf{x}\sim p}h(\mathbf{x}^+,\mathbf{x}) \label{eq:lepn}\\&&~~~~~~~~~~ - \frac{\tau^+}{\tau^-}\mathbb{E}_{\mathbf{x}^\prime\sim p^+}h(\mathbf{x}^+, \mathbf{x}^\prime)]\nonumber
\end{eqnarray}
Applying the result of Eq~\eqref{eq:pmin}, we have
\begin{eqnarray}
&&\frac{1}{\tau^-}\mathbb{E}_{\mathbf{x}\sim p}h(\mathbf{x}^+,\mathbf{x}) - \frac{\tau^+}{\tau^-}\mathbb{E}_{\mathbf{x}^\prime\sim p^+}h(\mathbf{x}^+, \mathbf{x}^\prime)\nonumber \\
&=& \int_\mathbf{x^-}h(\mathbf{x}^+,\mathbf{x}^-)p^-(\mathbf{x}^-)d\mathbf{x}^- \nonumber \\
&=& \mathbb{E}_{\mathbf{x}^-\sim p^-}h(\mathbf{x}^+,\mathbf{x}^-) \nonumber\\
&=&\mathbb{P}_\text{PN} \label{eq:epn}
\end{eqnarray}
Inserting Eq~\eqref{eq:epn} back to Eq~\eqref{eq:lepn} we obtain
\[\mathcal{L}_\textsc{dpl}  \rightarrow \mathcal{L}_\textsc{sup},\]
which completes the proof.
\end{proof}

Lemma~\ref{lemma:asy} demonstrates that the DPL loss $\mathcal{L}_\textsc{dpl}$ is asymptotically consistent with the idealized loss $\mathcal{L}_\textsc{sup}$ as M and N approach infinity. However, in practical applications, only a finite number of M and N are available, resulting in the empirical estimate $\hat{\mathcal{L}}\textsc{dpl}$. Next, Lemma~\ref{lemma:err} bounds the estimation error $|\hat{\mathcal{L}}_\textsc{dpl}-\mathcal{L}_\textsc{sup}|$.
\end{lemma}
\begin{lemma}\label{lemma:err}
With probability at
	least $1-\delta$, we have
	\begin{eqnarray}
	&&|\hat{\mathcal{L}}_\textsc{dpl}-\mathcal{L}_\textsc{sup}| \leq \nonumber e^2\sqrt{\frac{2\pi}{N}} + e^2\tau^+\sqrt{\frac{2\pi}{M}}
	\end{eqnarray}
	where $ \mathfrak{R}_n$ is bipartite Rademacher complexity.
\begin{proof}
Without loss of generality, we use the cosine similarity as score function to simplify the analysis, meaning that all embeddings are mapped on a hypersphere with radius 1.
Recall that $h$ is a function that maps two samples into a sigmoid function,$h(\mathbf{x}^+,\mathbf{x}) = \sigma (g(\mathbf{x}^+)- g(\mathbf{x}))$. As, $g(\cdot) \in [0,1]$, so  $-2 \leq g(\mathbf{x}^+)- g(\mathbf{x}) \leq 2$, and $ \frac{1}{1+e^{2}} \leq h(\mathbf{x}^+,\mathbf{x}) \leq \frac{1}{1+e^{-2}} $.

For fixed the positive sample $\mathbf{x}^+$, we denote the difference between the integrands of the asymptotic and non-asymptotic objective as $\triangle$:
\begin{eqnarray}
	\triangle &=& | \log [ \frac{1}{N\tau^-}  \sum_{\mathbf{x}} h(\mathbf{x}^+,\mathbf{x})  -\frac{\tau^+}{M\tau^-} \sum_{\mathbf{x}^\prime} p(\mathbf{x}^+,\mathbf{x}^\prime)] \nonumber\\
	&& - \log\mathbb{E}_{\mathbf{x}^-\sim p^-}h(\mathbf{x}^+,\mathbf{x}^-)|\nonumber \\
	&=&| \log [ \frac{1}{N\tau^-}  \sum_{\mathbf{x}} h(\mathbf{x}^+,\mathbf{x})  -\frac{\tau^+}{M\tau^-} \sum_{\mathbf{x}^\prime} p(\mathbf{x}^+,\mathbf{x}^\prime)] \nonumber\\
	&& - \log\mathbb{E}_{\substack{\mathbf x \sim p(\mathbf x) \\ \mathbf x^\prime \sim p^+(\mathbf x)}} [ \frac{1}{\tau^-}h(\mathbf{x}^+,\mathbf{x}) - \frac{\tau^+}{\tau^-}h(\mathbf{x}^+,\mathbf{x}^\prime)]| \nonumber\\
	&=& | \log \frac{\frac{1}{N}  \sum_{\mathbf{x}} h(\mathbf{x}^+,\mathbf{x})  -\frac{\tau^+}{M} \sum_{\mathbf{x}^\prime} h(\mathbf{x}^+,\mathbf{x}^\prime)}{\mathbb{E}_{\substack{\mathbf x \sim p(\mathbf x) \\ \mathbf x^\prime \sim p^+(\mathbf x)}} [ h(\mathbf{x}^+,\mathbf{x}) - \tau^+h(\mathbf{x}^+,\mathbf{x}^\prime)]} | \nonumber
\end{eqnarray}
 We first seek to bound the probability that $\triangle$ exceeds $\epsilon$ for fixed $\mathbf{x}^+$. Applying the fact $\mathbb{P}(|X|\geq \epsilon )=\mathbb{P}(X\geq \epsilon )+\mathbb{P}(-X\geq \epsilon )$, we have
\begin{eqnarray}
\mathbb{P}(\triangle \geq \epsilon) = \mathbf{I}(\epsilon) + \mathbf{II}(\epsilon) 
\end{eqnarray}
where
\begin{eqnarray}
&&\mathbf{I}(\epsilon) \nonumber \\
&=& \mathbb{P} \left(\log \frac{\frac{1}{N}  \sum_{\mathbf{x}} h(\mathbf{x}^+,\mathbf{x})  -\frac{\tau^+}{M} \sum_{\mathbf{x}^\prime} h(\mathbf{x}^+,\mathbf{x}^\prime)}{\mathbb{E}_{\substack{\mathbf x \sim p(\mathbf x) \\ \mathbf x^\prime \sim p^+(\mathbf x)}} [ h(\mathbf{x}^+,\mathbf{x}) - \tau^+h(\mathbf{x}^+,\mathbf{x}^\prime)]} \geq \epsilon  \right) \\
&\leq& \mathbb{P} \left( \frac{\frac{1}{N}  \sum_{\mathbf{x}} h(\mathbf{x}^+,\mathbf{x})  -\frac{\tau^+}{M} \sum_{\mathbf{x}^\prime} h(\mathbf{x}^+,\mathbf{x}^\prime)}{\mathbb{E}_{\substack{\mathbf x \sim p(\mathbf x) \\ \mathbf x^\prime \sim p^+(\mathbf x)}} [ h(\mathbf{x}^+,\mathbf{x}) - \tau^+h(\mathbf{x}^+,\mathbf{x}^\prime)]}-1 \geq \epsilon  \right) \label{eq:logx}\\
&\leq& \mathbb{P} ( \frac{1}{N}  \sum_{\mathbf{x}} h(\mathbf{x}^+,\mathbf{x})  -\frac{\tau^+}{M} \sum_{\mathbf{x}^\prime} h(\mathbf{x}^+,\mathbf{x}^\prime)  \nonumber \\ &&-\mathbb{E}_{\substack{\mathbf x \sim p(\mathbf x) \\ \mathbf x^\prime \sim p^+(\mathbf x)}} [ h(\mathbf{x}^+,\mathbf{x}) - \tau^+h(\mathbf{x}^+,\mathbf{x}^\prime)] \geq \frac{\epsilon}{1+e^2}  ) \label{eq:emin}
\end{eqnarray}
where Eq~\eqref{eq:logx} is obtained since $\log x \leq x-1$ for $x>0$. Eq~\eqref{eq:emin} is obtained since $\mathbb{E}_{\substack{\mathbf x \sim p(\mathbf x) \\ \mathbf x^\prime \sim p^+(\mathbf x)}} [ h(\mathbf{x}^+,\mathbf{x}) - \tau^+h(\mathbf{x}^+,\mathbf{x}^\prime)] =(1-\tau^+)\mathbb{E}_{\mathbf x \sim p^-(\mathbf x)}  h(\mathbf{x}^+,\mathbf{x}) \geq 1/(1+e^2)$. The second term is bounded similarly:
\begin{eqnarray}
&&\mathbf{II}(\epsilon) \nonumber\\
&=& \mathbb{P} \left(\log 
\frac{\mathbb{E}_{\substack{\mathbf x \sim p(\mathbf x) \\ \mathbf x^\prime \sim p^+(\mathbf x)}} [ h(\mathbf{x}^+,\mathbf{x}) - \tau^+h(\mathbf{x}^+,\mathbf{x}^\prime)]}{\frac{1}{N}  \sum_{\mathbf{x}} h(\mathbf{x}^+,\mathbf{x})  -\frac{\tau^+}{M} \sum_{\mathbf{x}^\prime} h(\mathbf{x}^+,\mathbf{x}^\prime)} \geq \epsilon  \right) \nonumber\\
&\leq& \mathbb{P} \left( \frac{\mathbb{E}_{\substack{\mathbf x \sim p(\mathbf x) \\ \mathbf x^\prime \sim p^+(\mathbf x)}} [ h(\mathbf{x}^+,\mathbf{x}) - \tau^+h(\mathbf{x}^+,\mathbf{x}^\prime)]}{\frac{1}{N}  \sum_{\mathbf{x}} h(\mathbf{x}^+,\mathbf{x})  -\frac{\tau^+}{M} \sum_{\mathbf{x}^\prime} h(\mathbf{x}^+,\mathbf{x}^\prime)} -1 \geq \epsilon  \right) \label{eq:logxii}\nonumber\\
&\leq& \mathbb{P} (\mathbb{E}_{\substack{\mathbf x \sim p(\mathbf x) \\ \mathbf x^\prime \sim p^+(\mathbf x)}} [ h(\mathbf{x}^+,\mathbf{x}) - \tau^+h(\mathbf{x}^+,\mathbf{x}^\prime)]  \nonumber \\ &&- [\frac{1}{N}  \sum_{\mathbf{x}} h(\mathbf{x}^+,\mathbf{x})  -\frac{\tau^+}{M} \sum_{\mathbf{x}^\prime} h(\mathbf{x}^+,\mathbf{x}^\prime)]  \geq \frac{\epsilon}{1+e^2}  ) \label{eq:eminii}
\end{eqnarray}
Combining Eq~\eqref{eq:emin} and Eq~\eqref{eq:eminii} we have 
\begin{eqnarray}
&&\mathbb{P}(\triangle \geq \epsilon) \nonumber \\
&\leq& \mathbb{P} ( |\frac{1}{N}  \sum_{\mathbf{x}} h(\mathbf{x}^+,\mathbf{x})  -\frac{\tau^+}{M} \sum_{\mathbf{x}^\prime} h(\mathbf{x}^+,\mathbf{x}^\prime)  \label{eq:abs}\\ &&-\mathbb{E}_{\substack{\mathbf x \sim p(\mathbf x) \\ \mathbf x^\prime \sim p^+(\mathbf x)}} [ h(\mathbf{x}^+,\mathbf{x}) - \tau^+h(\mathbf{x}^+,\mathbf{x}^\prime)]| \geq \frac{\epsilon}{1+e^2}  ) \nonumber\\
&=& \mathbb{P} (|[\frac{1}{N}  \sum_{\mathbf{x}} h(\mathbf{x}^+,\mathbf{x}) -\mathbb{E}_{\mathbf x \sim p(\mathbf x)}  h(\mathbf{x}^+,\mathbf{x}) ] \\
&&-[\frac{\tau^+}{M} \sum_{\mathbf{x}^\prime} h(\mathbf{x}^+,\mathbf{x}^\prime)  -\mathbb{E}_{\mathbf x^\prime \sim p^+(\mathbf x)}  \tau^+h(\mathbf{x}^+,\mathbf{x}^\prime)]| \geq \frac{\epsilon}{1+e^2}  ) \nonumber\\
&\leq& \mathbb{P} (|\frac{1}{N}  \sum_{\mathbf{x}} h(\mathbf{x}^+,\mathbf{x}) -\mathbb{E}_{\mathbf x \sim p(\mathbf x)}  h(\mathbf{x}^+,\mathbf{x})| \label{eq:abs1}\\
&&+|\frac{\tau^+}{M} \sum_{\mathbf{x}^\prime} h(\mathbf{x}^+,\mathbf{x}^\prime)  -\mathbb{E}_{\mathbf x^\prime \sim p^+(\mathbf x)}  \tau^+h(\mathbf{x}^+,\mathbf{x}^\prime)| \geq \frac{\epsilon}{1+e^2}  ) \nonumber\\
&\leq& \mathbf{III} (\epsilon) + \mathbf{IV} (\epsilon). \label{eq:abs2}
\end{eqnarray}
where 
\begin{eqnarray}
\mathbf{III} (\epsilon) &=& \mathbb{P}(|\frac{1}{N}  \sum_{\mathbf{x}} h(\mathbf{x}^+,\mathbf{x}) -\mathbb{E}_{\mathbf x \sim p(\mathbf x)}  h(\mathbf{x}^+,\mathbf{x})| \nonumber\\
&&\geq \frac{\epsilon}{2(1+e^2)}  ) \\
\mathbf{IV} (\epsilon) &=&\mathbb{P}(|\frac{\tau^+}{M} \sum_{\mathbf{x}^\prime} h(\mathbf{x}^+,\mathbf{x}^\prime)  -\mathbb{E}_{\mathbf x^\prime \sim p^+(\mathbf x)}  \tau^+h(\mathbf{x}^+,\mathbf{x}^\prime)|\nonumber\\
&&\geq \frac{\epsilon}{2(1+e^2)}  ) 
\end{eqnarray}
Eq~\eqref{eq:abs1} is obtained due to $|X-Y| \leq |X|+|Y|$, Eq~\eqref{eq:abs2} is obtained due to
$\mathbb{P}(|X|+|Y| \leq \epsilon) \leq \mathbb{P}(|X| \leq \epsilon/2) + \mathbb{P}(|Y| \leq \epsilon/2)$.
McDiarmid's inequality states that, for independent random variables $X_{1},X_{2},\dots ,X_{n}$, where ${\displaystyle X_{i}\in {\mathcal {X}}_{i}}$ for all $i$, if ${\displaystyle f:{\mathcal {X}}_{1}\times {\mathcal {X}}_{2}\times \cdots \times {\mathcal {X}}_{n}\rightarrow \mathbb {R} }$ satisfy the bounded differences property with bounds ${\displaystyle c_{1},c_{2},\dots ,c_{n}}$, then, for any $\epsilon >0$
\begin{eqnarray}
\mathbb{P}(|f(X_{1},X_{2},\ldots ,X_{n})-\mathbb {E} [f(X_{1},X_{2},\ldots ,X_{n})]|\geq \epsilon ) \nonumber\\  \leq 2\exp \left(-{\frac {2\epsilon ^{2}}{\sum _{i=1}^{n}c_{i}^{2}}}\right). \nonumber
\end{eqnarray}
In our particular case, let score of unlabeled sample $g(\mathbf{x})$ be random variable, since $ \frac{1}{1+e^{2}} \leq h(\mathbf{x}^+,\mathbf{x}) \leq \frac{1}{1+e^{-2}} $, the function $f:{\mathcal {X}}_{1}\times {\mathcal {X}}_{2}\times \cdots \times {\mathcal {X}}_{n}\rightarrow \frac{1}{N}  \sum_{\mathbf{x}}  h(\mathbf{x}^+,\mathbf{x})$ satisfies the bounded differences property with bounds ${\displaystyle c_{1}=c_{2}=\dots =c_{n}} = \frac{1}{N}\frac{e^2-1}{e^2+1}$, yielding the following bound:
\begin{eqnarray}
\mathbf{III}(\epsilon)  
&\leq& 2\exp \left(-\frac{N\epsilon^2}{2(e^2-1)^2} \right)  \\
&\leq& 2\exp \left(-\frac{N\epsilon^2}{2e^4} \right).\label{eq:p3}
\end{eqnarray}
Similarly, let score of positive sample $g(\mathbf{x}^\prime)$ be random variable, the function $f:{\mathcal {X}}_{1}\times {\mathcal {X}}_{2}\times \cdots \times {\mathcal {X}}_{m}\rightarrow \frac{\tau^+}{M} \sum_{\mathbf{x}}  h(\mathbf{x}^+,\mathbf{x}^\prime) $ satisfies the bounded differences property with bounds ${\displaystyle c_{1}=c_{2}=\dots =c_{m}} = \frac{\tau^+}{M}\frac{e^2-1}{e^2+1}$, yielding the following bound:
\begin{eqnarray}
\mathbf{IV}(\epsilon)  
&\leq& 2\exp \left(-\frac{M\epsilon^2}{2(e^2-1)^2{\tau^+}^2} \right)  \\
&\leq& 2\exp \left(-\frac{M\epsilon^2}{2e^4{\tau^+}^2}  \right).\label{eq:p4}
\end{eqnarray}
Inserting Eq~\eqref{eq:p3} and Eq~\eqref{eq:p4} back to Eq~\eqref{eq:abs2}, we have
\begin{eqnarray}
	\mathbb{P}(\triangle \geq \epsilon|\mathbf{x}^+) \leq 2\exp \left(-\frac{N\epsilon^2}{2e^4} \right)+2\exp \left(-\frac{M\epsilon^2}{2e^4{\tau^+}^2}  \right). \label{eq:tail}
\end{eqnarray}
Our objective is to bound the term $|\mathcal{L}_{\text{DPL}}(g) - \hat{\mathcal{L}}_\text{DPL}(g)| $, to do this we follow DCL~\cite{Chuang:2020:NIPS} to push the absolute value inside the expectation by Jensen's inequality
\begin{eqnarray}
&&|\hat{\mathcal{L}}_\textsc{dpl}-\mathcal{L}_\textsc{sup}|\nonumber\\
&=& \mathbb{E}_{\mathbf{x}^+} \log \mathbf{I} - \mathbf{II}| \\
 &\leq& \mathbb{E}_{\mathbf{x}^+} \triangle\\
&=&  \mathbb{E}_{\mathbf{x}^+} [\mathbb{E}_\epsilon[\triangle|\mathbf{x}^+]] \\
&=&  \mathbb{E}_{\mathbf{x}^+} \left[\int_{0}^{+\infty} \mathbb{P}(\triangle \geq \epsilon|\mathbf{x}^+)d\epsilon \right] \label{eq:tail1} \\
&\leq& \int_{0}^{+\infty}  2\exp \left(-\frac{N\epsilon^2}{2e^4} \right)+2\exp \left(-\frac{M\epsilon^2}{2e^4{\tau^+}^2}  \right) d \epsilon \nonumber\\
&=& e^2\sqrt{\frac{2\pi}{N}} + e^2\tau^+\sqrt{\frac{2\pi}{M}}.
\end{eqnarray}
The outer expectation in Eq.~\eqref{eq:tail1} disappears since the tail probably bound holds uniformly for
all fixed positive sample $\mathbf{x}^+$~\cite{Chuang:2020:NIPS}.
\end{proof}
\end{lemma}

\section{Experiment}
\subsection{Experiment Settings}
\subsubsection{Dataset}
We conduct our experiments on five publicly available datasets: MovieLens-100k, MovieLens-1M, Yahoo!-R3, Yelp2018 and Gowalla. These datasets comprise users' ratings on items using a discrete five-point grading system, providing information on the items that the users have interacted with. The first 3 datasets contain user ratings, we follow~\cite{Steffen:2009:UAI,Zhang:2013:SIGIR,Steffen:2014:WSDM} to convert all rated items to implicit feedback. For each dataset, we randomly allocate 20\% of the data as test data, with the remaining 80\% used for training. Table~\ref{Table:Dataset} presents a summary of the dataset statistics.
\begin{table*}[h!]
	\centering
	\small
	\caption{Dataset Statistics}\label{Table:Dataset}
	\begin{tabular}{lrrrrrr}
		\toprule[1.2pt]
		~           & users   & items  &interactions & training set  &test set&density  \\ \cline{1-7}
		MovieLens-100k   &   943    &  1,682   &100,000&    80k	   & 20k &0.06304\\
		MovieLens-1M    &   6,040  &  3,952   &1,000,000&  800k     & 200k&0.04189  \\
		Yahoo!-R3       &   5,400  &  1,000  &182,000 &   146k      & 36k&0.03370\\
		Yelp2018       &   31,668  &  38,048&   1,561,406&   1,249k     & 312k&0.00130  \\
		Gowalla       &   29,858 &  40,981  &1,027,370 &   821k     & 205k&0.00084 \\
		\bottomrule[1.2pt]
	\end{tabular}
\end{table*}
\subsubsection{Evaluation metric}
In order to evaluate the performance of the recommendations, we adopt commonly used metrics, namely precision (P), recall (R), and normalized discounted cumulative gain (NDCG), to assess the Top-$K$ recommendations, where $K$ is selected as 5, 10, and 20. For the sake of brevity, we assume that the definitions of these metrics are widely known and do not provide them here.
\subsubsection{Experimental Setup}
In our experimental setup, we consider two recommendation models: the classic \textit{matrix factorization} (MF)\cite{Koren:2009:Computer} and the more recent \textit{light graph convolution network} (LightGCN)\cite{Xiangnan:2020:SIGIR}. The computations for the first three datasets were conducted on a personal computer running the Windows 10 operating system with a 2.1 GHz CPU, an RTX 1080Ti GPU, and 32 GB of RAM. The computations for the last two datasets were performed on a cloud server running the Linux operating system with a Xeon(R) Platinum 8358P CPU, an RTX A40 GPU, and 56GB of RAM. The code and corresponding parameters have been released at: \url{https://github.com/liubin06/DPL} for reproducibility.
\subsubsection{Baselines}
\begin{itemize}
	\item BPR~\cite{Steffen:2009:UAI}: BPR introduces the pairwise learning approach based on maximum a posteriori estimation for implicit collaborative filtering. BPR and NCE have identical mathematical forms, but are described differently. Since implicit feedback only involves  positive data of interactions and unlabeled data, BPR samples positives  from the $(u,i)$  pairs that users interacted with, while the negative examples are drawn from the unlabeled data $(u,j)$ that users have not interacted with.
	\item InfoNCE~\cite{Oord:2018:arxiv}: 
	The InfoNCE loss is a popular loss function used in machine learning, particularly in the context of representation learning.  Specifically, InfoNCE measures the similarity between a query sample $\mathbf{x}^+$ and the set of negative samples $\{\mathbf{x}_i^-\}_{i=1}^N$  and applying a softmax function:
		\begin{eqnarray}\label{eq:Info}
		\mathcal{L}_\text{InfoNCE}  \nonumber
		&&=\\  - \mathbb{E}_{\substack{\mathbf x^+ \sim p^+(\mathbf x) \\ \mathbf x_i^- \sim p^-(\mathbf x)}}&&\log\frac{\exp(g(\mathbf{x}^+))}{\exp(g(\mathbf{x}^+))+ \sum_{i=1}^{N}\exp( g(\mathbf{x}_i^-))} \nonumber
	\end{eqnarray}
	InfoNCE can be seen as a generalization of Noise-Contrastive Estimation (NCE) from one negative sample to N negative samples. In practice, since label of negative samples are unavailable, $\mathbf{x}_i^-$ are typically  sampled from unlabeled samples.
	\item DCL~\cite{Chuang:2020:NIPS}: Due to the presence of false negatives in the unlabeled data, DCL corrects the probability estimates to perform false negative debiasing. Specifically, it proposes the estimator to replace the second term in the denominator of the $\mathcal{L}_\text{InfoNCE}$:
\begin{eqnarray}\label{eq:DCL}
	\mathcal{L}_\text{DCL} 
	&=&  - \mathbb{E}_{\substack{\mathbf x^+ \sim p^+(\mathbf x) \\ \mathbf x_i^- \sim p^-(\mathbf x)}}\log\frac{\exp(g(\mathbf{x}^+))}{\exp(g(\mathbf{x}^+))+ Ng} \nonumber
\end{eqnarray}
where 
\begin{eqnarray}\nonumber
g =  \frac{1}{N\tau^-}  (\sum_{i=1}^{N} \exp(g(\mathbf{x}_i) - N\tau^+ \cdot \frac{\sum_{j=1}^{K} \exp(g(\mathbf{x}^+_j)}{K} ) \label{Eq:DCLEstimator}
\end{eqnarray}
The $\text{g}$ estimator can be interpreted as the summation of scores of true negative samples. Specifically, $N\tau^+$ estimates the number of false negative samples, and $\frac{\sum_{j=1}^{K} \exp(g(\mathbf{x}^+_j)}{K} $ estimates the mean value of scores of $K$ false negative samples. Thus, the second term inside the parentheses corresponds to the summation of scores of all false negative samples among $N$ samples, while subtracting it from summation of N unlabeled scores $\sum_{i=1}^{N} \exp(g(\mathbf{x}_i)$ corresponds to the summation of all true negative samples among $N$ randomly selected  unlabeled samples.

	\item HCL~\cite{Robinson:2021:ICLR}: Following the DCL debiasing framework, it also takes into consideration of hard negative mining by up-weighting each randomly selected unlabeled sample as follows.
	\begin{eqnarray}\label{eq:hcl}
		\omega_i^\textsc{Hcl} = \frac{g(\mathbf{x}^+_j)^\beta}{\frac{1}{N} \sum_{j=1}^{N}g(\mathbf{x}^+_j)^\beta}.
	\end{eqnarray}
where beta controls the hardness level for mining hard negatives. DCL is a particular case of HCL with $\beta=0$.
\end{itemize}
\subsection{Experimental Results}
\subsubsection{Recommendation Performance}
\begin{table*}[h!]
	\centering
	\caption{Performance comparison of five learning algorithms on five datasets.}\label{Table:Recommendation}
	\resizebox{1\textwidth}{!}{
		\begin{tabular}{lllccccccccccc}
			\toprule[1.2pt]
			\multirow{2}*{\textbf{Dataset}} & \multirow{2}*{\textbf{CF Model}} & \multirow{2}*{\textbf{Learning Method}} & \multicolumn{3}{c}{Top-5} &~& \multicolumn{3}{c}{Top-10}&~&\multicolumn{3}{c}{Top-20}\\ \cline{4-6} \cline{8-10} \cline{12-14}
			~ & ~ & ~ & Precision& Recall& NDCG& ~ &Precision& Recall& NDCG& ~ &Precision& Recall& NDCG \\ \hline
			
			\multirow{12}*{\textbf{MovieLens-100k}} & \multirow{6}*{\textbf{MF}} & BPR & 0.3900   &0.1301	&0.4143	&~&0.3363	&0.2164	&0.3967& ~&0.2724&0.3298&0.3962 \\
			~ & ~ & InfoNCE  &0.4168 & 0.1434 & 0.4458 & ~ & 0.3513 & 0.2291 & 0.4202 & ~ & 0.2835 & 0.3546 & 0.4207 \\
			~ & ~ & DCL &0.4081 & 0.1388 & 0.4324 & ~ & 0.3452 & 0.2266 & 0.4095 & ~ & 0.2793 & 0.3497 & 0.4118 \\ 
			~ & ~ & HCL  & 0.4263 & 0.1463 & 0.4539 & ~ & 0.3565 & 0.2323 & 0.426 & ~ & 0.2849 & 0.3564 & 0.4242 \\	
			~ & ~ &DPL(Proposed)    &0.4348 & 0.1523 & 0.4643 & ~ & 0.3635 & 0.2379 & 0.4356 & ~ & 0.2914 & 0.3588 & 0.4338 \\
			\cline{2-14}
			
			~ & \multirow{6}*{\textbf{LightGCN}}  & BPR & 0.3944 & 0.1231 & 0.4204 & ~ & 0.3346 & 0.2189 & 0.4017 & ~ & 0.2658 & 0.3281 & 0.3986 \\ 
			~ & ~ & Info\_NCE & 0.3924 & 0.1343 & 0.4209 & ~ & 0.3349 & 0.2183 & 0.4006 & ~ & 0.2679 & 0.3289 & 0.3976 \\ 
			~ & ~ & DCL & 0.3962 & 0.1367 & 0.4243 & ~ & 0.3361 & 0.2194 & 0.4022 & ~ & 0.2695 & 0.3329 & 0.4006 \\ 
			~ & ~ & HCL & 0.4197 & 0.1461 & 0.4501 & ~ & 0.3458 & 0.2256 & 0.4188 & ~ & 0.2802 & 0.3446 & 0.4182 \\
			~ & ~ & DPL(proposed) & 0.4333 & 0.1486 & 0.4627 & ~ & 0.3596 & 0.2344 & 0.4324 & ~ & 0.2919 & 0.3585 & 0.4331 \\ \hline\hline

		\multirow{12}*{\textbf{MovieLens-1M}} & \multirow{6}*{\textbf{MF}} & BPR & 0.3929 & 0.0922 & 0.4142 & ~ & 0.3411 & 0.152 & 0.3836 & ~ & 0.2839 & 0.237 & 0.368 \\ 
		~ & ~ & InfoNCE & 0.4009 & 0.0934 & 0.4209 & ~ & 0.3472 & 0.1546 & 0.3894 & ~ & 0.289 & 0.2423 & 0.3731 \\ 
		~ & ~ & DCL & 0.3820 & 0.0879 & 0.4003 & ~ & 0.3339 & 0.1478 & 0.3728 & ~ & 0.2821 & 0.2358 & 0.3605 \\ 
		~ & ~ & HCL & 0.4112 & 0.0969 & 0.4317 & ~ & 0.3552 & 0.1585 & 0.3991 & ~ & 0.2959 & 0.2475 & 0.3825 \\ 
		~ & ~ & DPL(proposed) & 0.4212 & 0.0998 & 0.4407 & ~ & 0.3624 & 0.1625 & 0.4071 & ~ & 0.2991 & 0.2518 & 0.3891 \\ 
		\cline{2-14}
		~ & \multirow{6}*{\textbf{LightGCN}} & BPR & 0.3517 & 0.0739 & 0.3726 & ~ & 0.2997 & 0.1201 & 0.3385 & ~ & 0.2467 & 0.1884 & 0.3172 \\ 
		~ & ~ & InfoNCE & 0.4121 & 0.0986 & 0.4386 & ~ & 0.359 & 0.1594 & 0.4041 & ~ & 0.2979 & 0.2482 & 0.3869 \\ 
		~ & ~ & DCL & 0.4104 & 0.0982 & 0.4291 & ~ & 0.3544 & 0.1597 & 0.3977 & ~ & 0.2965 & 0.2511 & 0.3842 \\ 
		~ & ~ & HCL & 0.4107 & 0.0948 & 0.4300 & ~ & 0.3514 & 0.1542 & 0.3950 & ~ & 0.2916 & 0.2413 & 0.3775 \\ 
		~ & ~ & DPL(proposed) & 0.4217 & 0.1003 & 0.4429 & ~ & 0.3620 & 0.1625 & 0.1866 & ~ & 0.2989 & 0.2511& 0.3896 \\\hline \hline
		
		\multirow{12}*{\textbf{Yahoo!-R3}} & \multirow{6}*{\textbf{MF}} & BPR & 0.1417 & 0.1052 & 0.1587 & ~ & 0.1064 & 0.1573 & 0.1641 & ~ & 0.0768 & 0.2259 & 0.1913 \\ 
		~ & ~ & Info\_NCE & 0.1454 & 0.1083 & 0.1635 & ~ & 0.1091 & 0.1618 & 0.1692 & ~ & 0.079 & 0.2327 & 0.1974 \\ 
		~ & ~ & DCL & 0.1429 & 0.1065 & 0.1615 & ~ & 0.1080 & 0.1601 & 0.1664 & ~ & 0.0786 & 0.2316 & 0.1952 \\ 
		~ & ~ & HCL & 0.1460 & 0.1097 & 0.1638 & ~ & 0.1096 & 0.1628 & 0.1697 & ~ & 0.0792 & 0.2336 & 0.1976 \\ 
		~ & ~ & DPL(proposed) & 0.1491& 0.1091 & 0.1652 & ~ & 0.1108 & 0.1641 & 0.1712 & ~ & 0.0801 & 0.2351 & 0.2012 \\ 
		\cline{2-14}
		~ & \multirow{6}*{\textbf{LightGCN}} &BPR & 0.1115 & 0.0838 & 0.01252 & ~ & 0.0881 & 0.1322 & 0.1346 & ~ & 0.0661 & 0.1976 & 0.1611 \\ 
		~ & ~ & Info\_NCE & 0.1456 & 0.1092 & 0.1642 & ~ & 0.1089 & 0.1622 & 0.1697 & ~ & 0.079 & 0.2333 & 0.1982 \\ 
		~ & ~ & DCL & 0.1417 & 0.1074 & 0.1676 & ~ & 0.1099 & 0.1633 & 0.1719 & ~ & 0.0798 & 0.2354 & 0.2007 \\ 
		~ & ~ & HCL & 0.1412 & 0.1139 & 0.1718 & ~ & 0.113 & 0.1683 & 0.1776 & ~ & 0.0812 & 0.2394 & 0.2059 \\ 
		~ & ~ & DPL(proposed) & 0.1504 & 0.1111 &  0.1697 & ~ & 0.1131 & 0.1670  & 0.1757 & ~ & 0.0825 & 0.2412 & 0.2054\\ \hline\hline

		\multirow{12}*{\textbf{Yelp2018}} & \multirow{6}*{\textbf{MF}} & BPR & 0.0398 & 0.0228 & 0.0435 & ~ & 0.0339 & 0.0389 & 0.0456 & ~ & 0.0284 & 0.065 & 0.0538 \\ 
		~ & ~ & Info\_NCE  & 0.0429 & 0.0246 & 0.047 & ~ & 0.0365 & 0.0417 & 0.0491 & ~ & 0.0305 & 0.07 & 0.058 \\ 
		~ & ~ & DCL & 0.0486 & 0.0278 & 0.0531 & ~ & 0.041 & 0.0466 & 0.0552 & ~ & 0.0342 & 0.0777 & 0.0648 \\
		~ & ~ & HCL & 0.0535 & 0.0535 & 0.0586 & ~ & 0.0459 & 0.0541 & 0.0622 & ~ & 0.0383 & 0.0894& 0.0736 \\ 
		~ & ~ & DPL(proposed) & 0.0543 & 0.0325 & 0.0595 & ~ & 0.0463 & 0.0551 & 0.0630 & ~ & 0.0389 & 0.0914 & 0.0749 \\
		\cline{2-14}
		~ & \multirow{6}*{\textbf{LightGCN}} & BPR & 0.0556 & 0.0330 & 0.0610 & ~ & 0.0473 & 0.0560 & 0.0644 & ~ & 0.0391 & 0.0914 & 0.0757 \\ 
		~ & ~ & Info\_NCE & 0.0553 & 0.0329 & 0.0607 & ~ & 0.0473 & 0.0558 & 0.0642 & ~ & 0.0390 & 0.0911 & 0.0754 \\ 
		~ & ~ & DCL & 0.0559 & 0.0331 & 0.0612 & ~ & 0.0472 & 0.0557 & 0.0642 & ~ & 	0.0391 & 0.0914 & 0.0756 \\ 
		~ & ~ & HCL & 0.0563 & 0.0335 & 0.0617 & ~ & 0.0477 & 0.0564 & 0.0648 & ~ & 0.0393 & 0.0920 & 0.0760 \\  
		~ & ~ & DPL(proposed) & 0.0604 & 0.0364 & 0.0657 & ~ & 0.0513 & 0.0615 & 0.0696 & ~ & 0.0423 & 0.1003 & 0.0821 \\\hline\hline

		\multirow{12}*{\textbf{Gowalla}} & \multirow{6}*{\textbf{MF}} & BPR & 0.0728 & 0.0748 & 0.1000 & ~ & 0.0555 & 0.1116 & 0.1063 & ~ & 0.0414 & 0.1625 & 0.1209 \\ 
		~ & ~ & Info\_NCE & 0.0739 &0.0757 & 0.1016 & ~ & 0.0560 & 0.1122 & 0.1076 & ~ & 0.0422 & 0.1650 & 0.1230\\ 
		~ & ~ & DCL & 0.0746 & 0.0769 &0.1023 & ~ & 0.0568 & 0.1147 & 0.1088 & ~ & 0.0426 & 0.1664 & 0.1238 \\ 
		~ & ~ & HCL & 0.0755 & 0.0774 & 0.1035 & ~ & 0.0574 & 0.1151 & 0.1098 & ~ & 0.0432& 0.1693 & 0.1256 \\ 
		~ & ~ & DPL(proposed) & 0.0815 & 0.0827 & 0.1100 & ~ & 0.0628 & 0.1243& 0.1174 & ~ & 0.0473 & 0.1815 & 0.1340 \\
		\cline{2-14}
		~ & \multirow{6}*{\textbf{LightGCN}} & BPR & 0.0735 & 0.0753 & 0.1007 & ~ & 0.0560 & 0.1119 & 0.1069 & ~ & 0.0419 & 0.1641 & 0.1218 \\ 
		~ & ~ & Info\_NCE & 0.0743 & 0.0760 & 0.1022 & ~ & 0.0566 & 0.1132& 0.1084 & ~ & 0.0423 & 0.1649 & 0.1231\\ 
		~ & ~ & DCL & 0.0748 & 0.0763 & 0.1027 & ~ & 0.0569 & 0.1132 & 0.1088 & ~ & 0.0424 & 0.1656 & 0.1236 \\ 
		~ & ~ & HCL & 0.0794 & 0.0804 & 0.1084 & ~ & 0.0608 &0.1199 & 0.1147 & ~ & 0.0453 & 0.1740 & 0.1319 \\ 
		~ & ~ & DPL(proposed) & 0.0867 &  0.0891 & 0.1164 & ~ &  0.0662 &  0.1329 & 0.1242 & ~ & 0.0494 & 0.1936 & 0.1417  \\ \hline
		\bottomrule[1.5pt]
			
		\end{tabular}
	}
\end{table*}
The first observation from Table~\ref{Table:Recommendation} is that DPL achieves the best performance. Compared to BPR and InfoNCE without debiasing mechanisms, DCL shows significant improvements, indicating the necessity of correcting biased probability estimates in implicit feedback data. Compared to DCL and HCL with debiasing mechanisms, DPL also achieves significant improvements, mainly due to the advantage of DPL's debiasing mechanism in the pairwise learning problem setting. In the pairwise learning problem setting, where the number of unlabeled samples is one and the true labels of the pairwise data corresponding to user preference can be enumerated (see Fig~\ref{fig:event}), an unbiased estimate of the probability that the user prefers positive examples over negative examples can be obtained. However, in the case of N unlabeled samples, according to the binomial theorem, there are $2^N$ possible outcomes for the true label, making it difficult to enumerate every case. Therefore, the debiasing mechanisms of DCL and HCL mainly rely on numerical approximation.

The second observation is that as a special case of InfoNCE with a negative sample size of 1, the BPR loss has a performance slightly inferior to that of InfoNCE with N negative samples, especially on large and sparse datasets. This is because a larger negative sample size N tighter lower bounds the mutual information~\cite{Oord:2018:arxiv}. However, in PU datasets, a larger N is not always better because a larger N usually result in larger gradient values to hard samples, i.e., samples that are embedded closer to the anchor point (i.e., user embedding) in the embedding space. If such a sample is a false negative, the larger gradient value can seriously damage the model's performance.

The third observation is that DCL, HCL, and DPL, which incorporate debiasing mechanisms, generally outperform BPR and InfoNCE, which lack debiasing mechanisms, highlighting the importance of debiasing on PU datasets, especially on datasets with a high positive class prior. Moreover, HCL assigns larger gradient values to negative samples by assigning higher weights to hard samples with high scores, achieving good results through implicit hard negative mining on top of debiasing. However, the corresponding hard negative mining parameter should be carefully tuned to prevent false negatives from harming model performance. This is because model performance will be harmed if an unlabeled hard sample is a false negative, but model performance will benefit from the hard sample if it is a true negative. This phenomenon is referred to as the "exploration-and-exploitation trade-off"\cite{Bin:2023:ICDE} in collaborative filtering and the "uniformity-alignment dilemma" in computer vision\cite{Feng:2021:CVPR}. 
\subsubsection{Hyperparameter Analysis}
\begin{figure*}[h!]
	\centering
	\includegraphics[width=\textwidth]{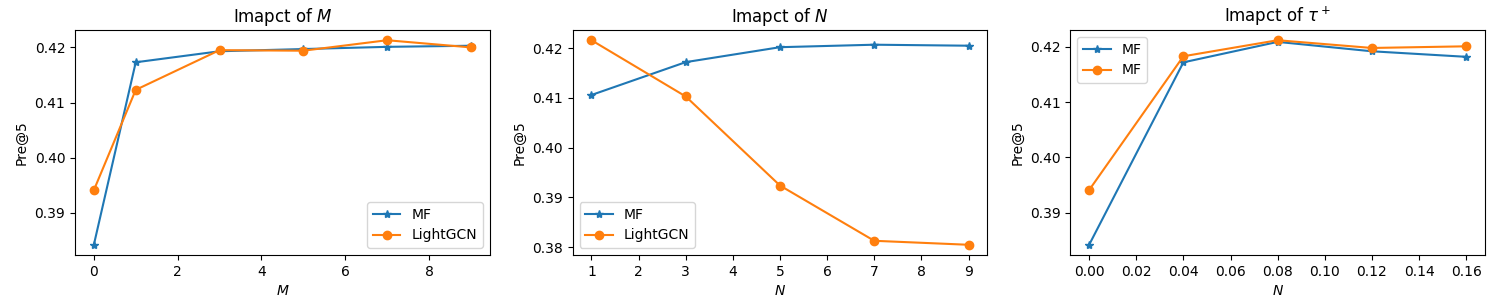}
	\caption{The changes of the Precision@5 metric under different parameters on the MovieLens-1M dataset. Generally, larger M leads to improved performance. However, different situations occur for parameter N. A particular note is that for matrix factorization models, a larger N should be set; for lightGCN models, a smaller N value should be set, such as N=1.} 
	\label{Fig:parameter}
\end{figure*}
Impact of M: The parameter M controls the number of additional positive examples used to correct the sampling bias. When M = 0, there is no debiasing mechanism. For the MF model, larger values of M and N usually lead to improved performance, with a significant performance boost observed from M=0 to M=1, highlighting the importance of using additional positive examples for debiasing. However, when M and N exceed a small constant, performance no longer improves. This is because the marginal gains in estimation accuracy from further increasing M and N become limited, as the theoretical analysis presented earlier.

Impact of N: The parameter N controls the number of negative examples used to compute the PU probability
Similar to the MF model, increasing M consistently improves LightGCN performance. However, as the number of negative examples N increases, the model performance unexpectedly decreases. We attribute this non-intuitive result to the fact that a larger N reduces the gradient value of hard negative samples. As Jensen's inequality states, $-\log(\frac{1}{N} \sum_{i=1}^N P_n) \leq \frac{1}{N} \sum_{i=1}^N \log P_n $; thus, the excessive value of N weakens the contribution of hard negative samples to the learning algorithm. The theoretical analysis assumes that the scores $g(\cdot)$ are independent and identically distributed (i.i.d.) variables, and the estimation error decreases as N increases. However, the aggregation mechanism of the graph neural network-based encoder seriously affects the i.i.d. property of $g(\cdot)$ values. Therefore, we recommend setting a small value of N for the LightGCN model.

Impact of positive class prior $\tau^+$: As $\tau^+$ increases, the model performance of MF and LightGCN exhibits an inverted U-shaped curve with an initial increase followed by a decrease. This is because setting a $\tau^+$ value that is too low or too high can lead to biased estimates of Formula 1. A common method to set the $\tau^+$ value is to treat the number of observed positive interactions $|\mathcal{D}^+|$ as a result of a Bernoulli trial, which occurs a total of $|\mathcal{U}|\times |\mathcal{I}|$ times, and succeeds in the number of interactions observed. The density $|\mathcal{D}^+|/(|\mathcal{U}|\times |\mathcal{I}|)$ of the dataset can then serve as a reference for setting the $\tau^+$ value. However, it should be noted that the $\tau^+$ value set based on the dataset density is a biased estimate that underestimates the true value, as all unobserved (u, i) pairs in the data are treated as negative samples, leading to an underestimation of the number of successes. We refer to~\cite{Jain:2016:NIPS,Christoffel:2016:ACML} for detailed discussion.

\subsubsection{DPL VS Negative Sampling}
Negative sampling and loss correction represent two distinct technical approaches in addressing the problem of sampling bias. The core idea of negative sampling is to select hard negative samples and use them for model training, which has demonstrated promising results. From a Bayesian statistical perspective, negative sampling utilizes two types of information: prior information, such as item category and popularity, which is static and used for sampling negative samples that users do not prefer; and sample information, such as scores and ranking positions, which is dynamic and continuously adjusted during model training, and used for sampling hard samples that are embedded close to the anchor embedding (higher scored). The differences between various negative sampling algorithms lie in how they utilize and process these two types of information. The latest Bayesian negative sampling algorithm specifies the negative signal measure in terms of posterior probability and proposes the theoretically optimal sampling rule, which has achieved good results.We compare the performance of DPL and BNS in Table~\ref{Table:vsns}.
\begin{table*}[h!]
	\centering
	\caption{Performance comparison of five learning algorithms on five datasets.}\label{Table:vsns}
	\resizebox{1\textwidth}{!}{
		\begin{tabular}{lllccccccccccc}
			\toprule[1.2pt]
			\multirow{2}*{\textbf{Dataset}} & \multirow{2}*{\textbf{CF Model}} & \multirow{2}*{\textbf{Method}} & \multicolumn{3}{c}{Top-5} &~& \multicolumn{3}{c}{Top-10}&~&\multicolumn{3}{c}{Top-20}\\ \cline{4-6} \cline{8-10} \cline{12-14}
			~ & ~ & ~ & Precision& Recall& NDCG& ~ &Precision& Recall& NDCG& ~ &Precision& Recall& NDCG \\ \hline
			
			\multirow{4}*{\textbf{MovieLens-100k}} & \multirow{4}*{\textbf{MF}} & BPR & 0.3900   &0.1301	&0.4143	&~&0.3363	&0.2164	&0.3967& ~&0.2724&0.3298&0.3962 \\
			~ & ~ &BNS   &0.4205 &0.1467	&0.4558&~	&0.3463	&0.2290	&0.4217& ~&0.2762&0.3466& 0.4176\\	
			~ & ~ &DPL(Proposed)    &0.4348 & 0.1523 & 0.4643 & ~ & 0.3635 & 0.2379 & 0.4356 & ~ & 0.2914 & 0.3588 & 0.4338 \\
			~ & ~ &DPL with Hard Samples   &0.4401 & 0.1579 & 0.4692 & ~ & 0.3713 & 0.2407 & 0.4395 & ~ & 0.2940 & 0.3592 & 0.4351 \\\hline \hline

			\multirow{4}*{\textbf{MovieLens-1M}} & \multirow{4}*{\textbf{MF}} & BPR & 0.3929 & 0.0922 & 0.4142 & ~ & 0.3411 & 0.152 & 0.3836 & ~ & 0.2839 & 0.237 & 0.368 \\ 
		    ~ & ~ & BNS&0.4207	&0.1062	&0.4324&~	&0.3518	&0.1703	&0.4191& ~&0.3045&0.2614&0.4002 \\
			~ & ~ & DPL(proposed) & 0.4212 & 0.0998 & 0.4407 & ~ & 0.3624 & 0.1625 & 0.4071 & ~ & 0.2991 & 0.2518 & 0.3891  \\
			~ & ~ & DPL with Hard Samples & 0.4251 & 0.1012 & 0.4412 & ~ & 0.3649 & 0.1701 & 0.4151 & ~ & 0.3012 & 0.2539 & 0.3922  \\\hline 
			\bottomrule[1.5pt]
		\end{tabular}
	}
\end{table*}
\begin{figure}[h!]
	\centering
	\includegraphics[width=0.5\textwidth]{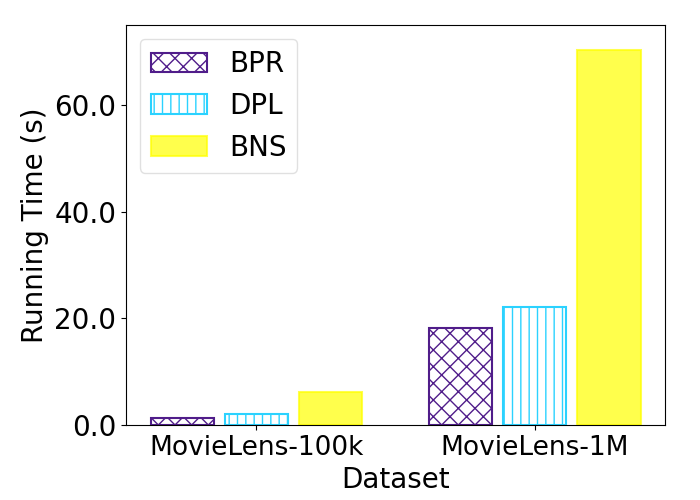}
	\caption{Comparison of running time. The running time is tested on a personal computer with a 2.1 GHz CPU, an RTX 1080Ti GPU, and 32 GB of RAM.} 
	\label{Fig:runtime}
\end{figure}

\textit{Performance}:
DPL achieved better performance on the MovieLens100k dataset and performed similarly to BNS on the MovieLens1M dataset, demonstrating the feasibility of using DPL as an alternative negative sampling algorithm based on correction estimation. Furthermore, we found that training DPL with hard, high-scored unlabeled samples can lead to some performance improvements. When training DPL with difficult samples, a relatively large tao value needs to be set. In Table~\ref{Table:vsns}, $\tau^+$ were set to 0.3 and 0.25, respectively, which are much higher than the density of the dataset itself. This is because training DPL with hard samples increases the probability of the model encountering false negative samples, which is equivalent to artificially changing the  positive class prior of training samples fed into the model.

\textit{Running Time}: Figure~\ref{Fig:runtime} displays the running time of one epoch training for the three algorithms on five datasets, with the CF model fixed as MF and the batch size fixed as 1024. BNS involves computing the empirical distribution function, and we implemented the in-batch approximation to save computational costs when computing the empirical distribution function. As shown in Figure~\ref{Fig:runtime}, the actual running time of DPL is only slightly longer than that of BPR, which is consistent with the time complexity analysis presented earlier. Even with the in-batch sample approximation to save computational costs, the running time of BNS is still 3-5 times longer than that of BPR and DPL. This is because dynamic negative sampling requires predicted scores to guide negative sampling, but GPU-based batch computation requires fixed negative samples before performing forward propagation to predict scores. As a result, dynamic negative sampling is typically implemented by loading additional negative samples as candidates into mini-batch data, leading to additional computational costs. Additionally, some state-of-the-art dynamic negative sampling algorithms require the ranking position of samples or the variance of predicted scores~\cite{Ding:2019:NeurIPS} in the previous training epochs, which requires the model to compute the predicted score of the entire user-item rating matrix rather than just the score within the mini-batch data when performing forward propagation, resulting in exponential time complexity. In summary, in scenarios with rich side information, we recommend using negative sampling algorithms that can flexibly combine prior information and model information. In scenarios where there is no available side information for supervision, we recommend using the DPL method for loss correction.

\section{Related Work}
\subsection{Collaborative Filtering}
Collaborative filtering was first formalized as a matrix completion problem for predicting scores~\cite{Koren:2009:Computer}. However, users typically only provide positive feedback by indicating their preferences or interests through interactions, resulting in binary values of 0 or 1 in the interaction matrix. BPR~\cite{Steffen:2009:UAI} introduced pairwise learning from pairwise comparisons of positive and negative item pairs to predict rankings. The core idea is to optimize the model to score the positive item higher than the negative item, which is reflected in the embedding space by pulling the positive item closer to the user and pushing the negative item further away. Mathematically, BPR and NCE~\cite{Gutmann:2010:ICAIS} are equivalent in their formulation~\cite{Liu:2021:TKDE}, but they are interpreted differently. Based on BPR optimization criterion, a series of recommendation models have been proposed, such as NGCF~\cite{Wang:2019:SIGIR} and LightGCN~\cite{Xiangnan:2020:SIGIR}, which have achieved state-of-the-art performance. Inspired by the success in CV and NLP, InfoNCE~\cite{Oord:2018:arxiv} has also been widely applied in collaborative filtering, which can be viewed as a generalization of BPR from one negative sample to N negative samples. Wu et al.\cite{wu2022effectiveness} prove that optimizing the InfoNCE loss is consistent with maximizing the Discounted Cumulative Gain (DCG) metric. Both BPR and InfoNCE face the problem of false negatives in unlabeled data, resulting in biased user-item representations\cite{Ding:2020:NIPS,Park:2019:WWW,Huang:2021:KDD,Ding:2019:IJCAI,Yang:2020:KDD}.

To address this issue, negative sampling algorithms have been extensively studied, such as graph-based negative sampling~\cite{Wang:2020:WWW, Chen:2019:WWW, Wang:2021:TKDE, Ying:2018:KDD}, and side or prior information-based negative sampling~\cite{Yuan:2016:IJCAI, Liu:2019:IJCAI, Jingtao:2019:IJCAI, Jingtao:2018:WWW}. We group them into two categories. The first kind is static negative sampling~\cite{Steffen:2009:UAI, Chen:2017:KDD, Mikolov:2013:NIPS, Weike:2013:IJCAI, Wang:2019:SIGIR}, which adopts a fixed sampling distribution, such as uniform sampling. The second kind is dynamic negative sampling~\cite{Steffen:2014:WSDM, Zhang:2013:SIGIR, Wang:2020:WWW, Chen:2019:WWW}. Algorithms of this kind favor negative instances with representations more similar to those of positive instances in the embedding space, for example, by selecting higher scored or higher ranked instances~\cite{Steffen:2014:WSDM, Zhang:2013:SIGIR}. However, they are more likely to suffer from the negative problem~\cite{Ding:2020:NIPS, Qin:2021:AAAI, Zhao:2021:IJCAI}. A novel class of methods generates virtual hard negatives from multiple unlabeled instances. These methods can be viewed as a generalization of the InfoNCE loss since the scores of the synthesized virtual negative samples are a function of the embeddings of unlabeled samples. For example, Huang et al.\cite{Huang:2021:KDD} propose to synthesize virtual hard negatives by hop mixing embeddings. Jun et al.\cite{Jun:2017:SIGIR} and Park et al.~\cite{Park:2019:WWW} design generative adversarial neural networks to generate virtual hard negatives.

\subsection{Contrastive Learning}
Contrastive learning is based on the "learn-to-compare" paradigm~\cite{Gutmann:2010:ICAIS,Steffen:2009:UAI}, which discriminates between positive and negative samples to avoid reconstructing pixel-level information of data~\cite{Oord:2018:arxiv}. While the representation encoder $f$ and similarity measure may vary across different domains, such as collaborative filtering~\cite{Steffen:2009:UAI,Xiangnan:2020:SIGIR,Wang:2019:SIGIR} and computer vision tasks~\cite{Devlin:2018:bert,He:2020:CVPR,Dosovitskiy:2014:NIPS}, they share the common idea of pulling positive samples closer to the anchor point while pushing the negative samples apart to train $f$ by optimizing a contrastive loss~\cite{Wang:2020:ICML}, such as BPR loss~\cite{Steffen:2009:UAI}, NCE loss~\cite{Gutmann:2010:ICAIS}, InfoNCE loss~\cite{Oord:2018:arxiv}, Infomax loss~\cite{Hjelm:2018:Arxiv}, asymptotic contrastive loss~\cite{Wang:2020:ICML}, among others. Supervised contrastive learning has achieved remarkable success in various domains~\cite{Henaff:2020:ICML,Khosla:2020:NIPS}, but it heavily relies on manually labeled datasets~\cite{Liu:2021:TKDE}. Self-supervised contrastive learning~\cite{Chen:2020:ICML,Chen:2020:NIPS,He:2020:CVPR,Henaff:2020:ICML,Xu:2022:Arxiv} has been extensively studied for its advantage in learning representations without requiring supervised data and has been shown to benefit a wide range of downstream tasks~\cite{Liu:2021:TKDE,Bachman:2019:NIPS,chen2020improved,Huang:2019:ICML,Wu:2018:CVPR,Zhuang:2019:CVPR}. In self-supervised contrastive learning, positive samples $x^+$ are obtained by applying a semantic-invariant operation on an anchor $x$ with heavy data augmentation, while negative samples $x^-$ are drawn from unlabeled data, which introduces the false negative problem and can lead to incorrect encoder training. This problem is related to the classic positive-unlabeled (PU) learning~\cite{Jessa:2020:ML,Du:2015:ICML,Du:2014:NIPS,Kiryo:2017:NIPS}. Existing empirical risk rewriting based methods for point-wise losses cannot be directly applied to contrastive loss. In recent years, several estimators that are consistent with supervised contrastive loss have been proposed, such as DCL~\cite{Chuang:2020:NIPS}, HCL~\cite{Robinson:2021:ICLR}, and BCL~\cite{Bin:2023:arxiv}. In particular, BCL proposes a posterior probability estimation of unlabeled samples being true negatives.

\section{Conclusion}
In this paper, we focus on addressing the problem of sampling bias from positive-unlabeled implicit feedback data,  but we adopt a different technical approach from explicit negative sampling. Specifically, we propose a correction for sampling bias from implicit feedback that yields a modified loss for pairwise learning called debiased pairwise loss (DPL). The key idea underlying DPL is to correct the biased probability estimates that result from false negatives, thereby correcting the gradients to approximate those of fully supervised data. The proposed objective is easy to implement and does not require additional side information for supervision or excessive storage and computational overhead.  In our future work, we will further explore the design of hard negative mining mechanisms on top of debiased pairwise loss.


\normalem
\bibliographystyle{IEEEtran}
\bibliography{ref}
\end{document}